\documentclass[journal=jacsat,manuscript=article]{achemso}

\usepackage{amsfonts}
\usepackage{verbatim}
\usepackage{amsmath}
\usepackage{amsthm}
\usepackage{amssymb}

\usepackage{xcolor}

\SectionNumbersOn
\author{Domenico Truzzolillo}
\altaffiliation{These authors contributed equally to this work}
\affiliation[Laboratoire Charles Coulomb (L2C), UMR 5221 CNRS-Universit\'{e} de Montpellier,
4 F-34095 Montpellier - France]{Laboratoire Charles Coulomb - Universit\'{e} de Montpellier - France}
\email{domenico.truzzolillo@umontpellier.fr}
\author{Simona Sennato}
\altaffiliation{These authors contributed equally to this work}
\affiliation[CNR-ISC UOS Roma- c/o Dipartimento di Fisica - Sapienza Universit\`{a} di Roma -
P.zzle A. Moro, 2 - 00185 Roma - Italy]{CNR-ISC UOS Sapienza, Roma - Italy}
\author{Stefano Sarti}
\affiliation[Dipartimento di Fisica - Sapienza Universit\`{a} di Roma - P.zzle A. Moro, 2 - 00185 Roma - Italy]{Dipartimento di Fisica - Sapienza Universit\`{a} di Roma - Italy}
\author{Stefano Casciardi}
\affiliation[National Institute for Insurance against Accidents at Work (INAIL Research), Department of Occupational and Environmental Medicine, Epidemiology and Hygiene, Roma - Italy]{INAIL Research - Roma - Italy}
\author{Chiara Bazzoni}
\affiliation[Dipartimento di Fisica - Sapienza Universit\`{a} di Roma - P.zzle A. Moro, 2 - 00185 Roma - Italy]{Dipartimento di Fisica - Sapienza Universit\`{a} di Roma - Italy}
\author{Federico Bordi}
\affiliation[CNR-ISC UOS Roma- c/o Dipartimento di Fisica - Sapienza Universit\`{a} di Roma -
P.zzle A. Moro, 2 - 00185 Roma - Italy]{CNR-ISC UOS Sapienza, Roma - Italy}
\alsoaffiliation[Dipartimento di Fisica - Sapienza Universit\`{a} di Roma - P.zzle A. Moro, 2 - 00185 Roma - Italy]{Dipartimento di Fisica - Sapienza Universit\`{a} di Roma - Italy}
\email{federico.bordi@roma1.infn.it}

\makeatletter
\let\acs@address@list\relax
\makeatother


\title
  {Overcharging and reentrant condensation of thermoresponsive ionic microgels}

\begin{document}
\setlength{\fboxrule}{0 pt}
\begin{tocentry}
\begin{LARGE}
\end{LARGE}
\vspace{0.5cm}
\begin{center}
\includegraphics[width=15cm]{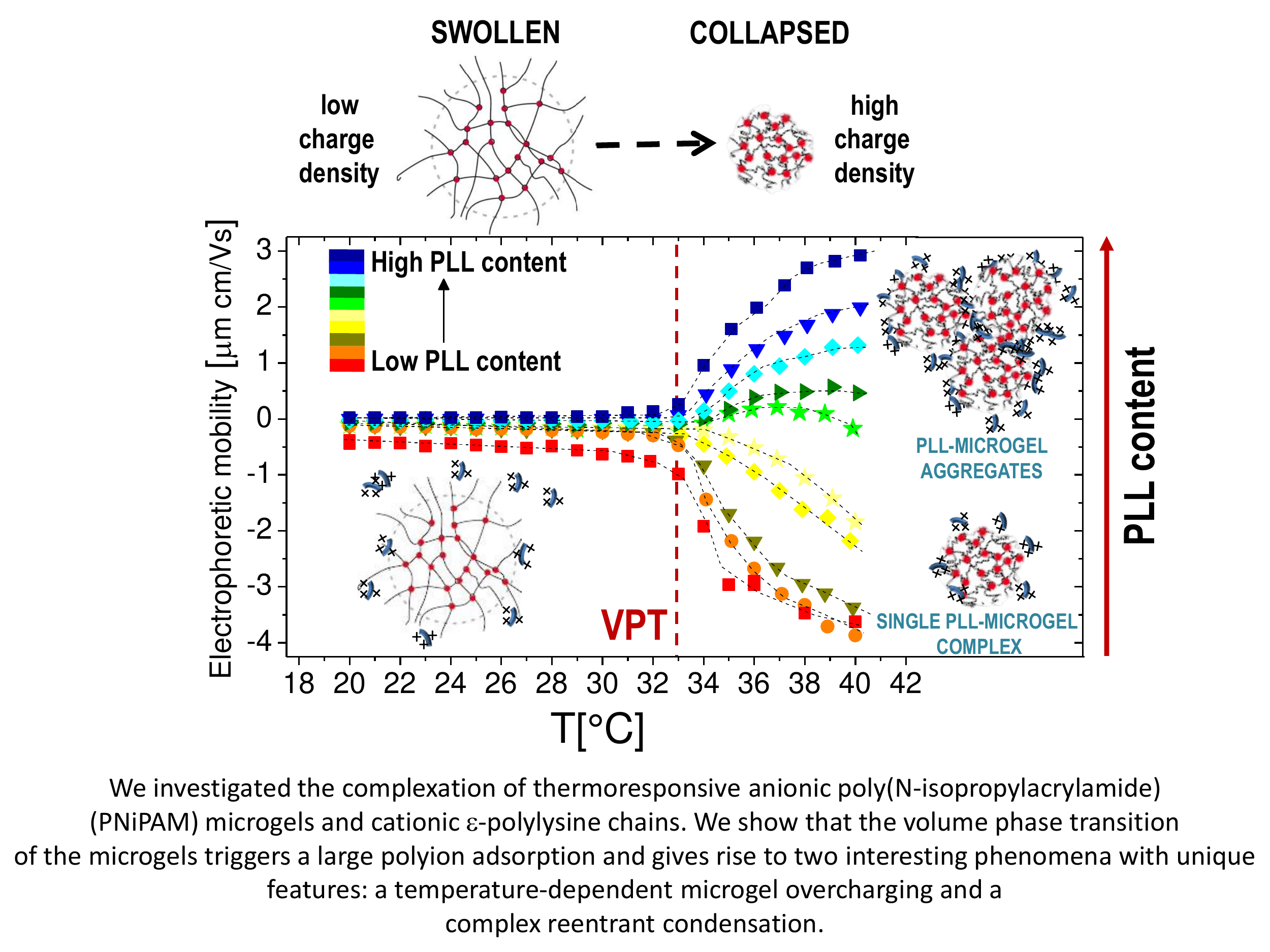}
\end{center}
\vspace{0.5cm}
\end{tocentry}
\begin{abstract}
We investigated the complexation of thermoresponsive anionic poly(N-isopropylacrylamide) (PNiPAM) microgels and cationic $\epsilon$-polylysine ($\epsilon$-PLL) chains. By combining electrophoresis, light scattering, transmission electron microscopy (TEM) and dielectric spectroscopy (DS) we studied the adsorption of $\epsilon$-PLL onto the microgel networks and its effect on the stability of the suspensions. We show that the volume phase transition (VPT) of the microgels triggers a large polyion adsorption. Two interesting phenomena with unique features occur: a temperature-dependent microgel overcharging and a complex reentrant condensation. The latter may occur at fixed polyion concentration, when temperature is raised above the VPT of microgels, or by increasing the number density of polycations at fixed temperature. TEM and DS measurements unambiguously show that short PLL chains adsorb onto microgels and act as electrostatic glue above the VPT. By performing thermal cycles, we further show that polyion-induced clustering is a quasi-reversible process: within the time of our experiments large clusters form above the VPT and partially re-dissolve as the mixtures are cooled down. Finally we give a proof that the observed phenomenology is purely electrostatic in nature: an increase of the ionic strength gives rise to the polyion desorption from the microgel outer shell.
\end{abstract}

\section{Introduction}
In aqueous solutions, oppositely charged colloids and polyelectrolytes, due to electrostatic interactions, self-assemble in complex aggregates \cite{bordi_polyelectrolyte-induced_2009}. The aggregation process may change considerably when different parameters, such as composition, ionic strength or colloid/polymer relative size are modified, and also due to different preparation procedures, exhibiting a rich and interesting phenomenology. The resulting supramolecular structures show quite different features on a mesoscopic scale, ranging from the 'neck-lace' case, where a long polyion chain ties together several particles in a 'beads and strings'-like manner \cite{radler_structure_1998}, to the case where several short polyions get adsorbed and 'decorate' the surface of each colloid\cite{sennato_decorated_2012}.

The complete understanding of the mechanism driving colloid-polyelectrolyte complexation still represents a fundamental problem of great interest in soft matter. Polyelectrolyte adsorption onto oppositely charged surfaces represents the core of this problem and a number of theoretical studies, using different approaches, have been published on this subject \cite{sens_counterion_2000,muthukumar_adsorption_1987,dobrynin_adsorption_2000}.

During the past few decades, colloid-multivalent ion complexation has been investigated by using either model systems, such as solid hard colloids \cite{bouyer_aggregation_2001}, soft colloids of biological interests \cite{bordi_polyelectrolyte-induced_2009}, or hydrophilic globular proteins \cite{zhang_reentrant_2008}. In all cases two distinct but intimately  related phenomena accompany and drive the self-assembly, i.e. charge-inversion and reentrant condensation.

Charge inversion occurs when, on a charged colloid particle, the total number of charges contributed by the (oppositely charged) adsorbed multivalent ions, crowding the surface, exceeds the original or 'bare' charge of the particle. As a consequence, the sign of the net charge of the resulting complex is opposite to that of the bare particle. Charge inversion originates by the strong lateral correlation between the adsorbed polyions \cite{grosberg_colloquium:_2002}, which generates a more or less ordered distribution of domains with excess negative charge (polyelectrolyte domains, in our case) and excess positive charge (polyelectrolyte-free domains). Indeed, by avoiding each other and residing as far away as possible to minimize their electrostatic interactions, adsorbed polyions leave the particle surface partially uncovered.
Such a non-homogeneous surface charge distribution, for systems where the long range electrostatic tails are sufficiently screened, originates a short range attractive interaction between the  so 'decorated' particles ('charge patch' attraction) \cite{velegol_analytical_2001, truzzolillo_interaction_2010,bordi_polyelectrolyte-induced_2009}.

Although these phenomena have been observed in a variety of polyelectrolyte-colloid mixtures in different conditions, in all previously reported works the charge density on the colloid surface was fixed, or, at least, it could not be changed without changing the ionic strength or the pH of the suspending medium. The charged thermoresponsive colloid considered in this work, being characterized by a thermodynamic volume phase transition (VPT), gives an opportunity of finely tuning the adsorption of polyelectrolytes simply by changing temperature. In fact, by changing the particle volume, VPT affects dramatically the charge density and hence the polyelectrolyte adsorption.

Poly(N-isopropylacrylamide) (PNiPAM) is a well-known thermosensitive microgel system, which exhibits a significant volume phase transition above the lower critical solution temperature (LCST), around $33$ $^\circ$C in aqueous media\cite{heskins_solution_1968,wu_volume_1996}. Therefore, this critical temperature is also called the volume phase transition temperature (VPTT).

The VPT of PNiPAM microgels has been extensively investigated,\cite{wu_volume_1996,tanaka_collapse_1978,kojima_cooperative_2010} not only because of its significant implications in a number of living phenomena, especially the protein folding and DNA packing\cite{kokufuta_effects_1993,sagle_investigating_2009}, but also due to the strong application background of this system, which is related to  the important feature of PNiPAM microgels that contain both hydrophilic amide groups and hydrophobic hydrocarbon chains.

It is well-known that the volume phase transition is determined by the hydrophobic interactions within the PNiPAM molecule. Indeed, many studies have shown that the VPTT is modified by the addition of inorganic salts,\cite{zhang_specific_2005} surfactants,\cite{kokufuta_effects_1993,richter_effect_2014} ionic liquids,\cite{reddy_interactions_2014,chang_influence_2015} alcohols,\cite{zhang_water/methanol_2001} and urea\cite{sagle_investigating_2009,gao_effect_2014}. Besides that, the VPTT and swelling/deswelling behavior are also modified by the introduction of charged groups (e.g., carboxyl, sulfonic, and amino group) into the PNiPAM microgel network.

For neutral PNiPAM microgels, the VPTT is mainly determined by two competing interactions, i.e. hydrogen bonding and hydrophobic interactions \cite{ilmain_volume_1991}, while for charged microgels, besides electrostatic effects, there is also an extra osmotic pressure contributing to their swelling, which arises from the ion/solvent mixing \cite{su_influence_2016}.
Therefore, though the volume phase transition of charged microgel is generally a more complex phenomenon to  be considered, nevertheless it offers the opportunity to tune charge density and penetrability just by changing the temperature, that are interesting features for a model system and very  appealing ones for biotechnological applications.

In this work we exploit the unique features of negatively  charged PNiPAM microgels to study their complexation with $\epsilon$-polylysine ($\epsilon$-PLL), a short cationic bio-compatible polymer.
We employed a combination of light scattering, electrophoretic and dielectric spectroscopy measurements to characterize $\epsilon$-PLL/PNiPAM complexes. We show that complexation is driven by the VPT of microgels, and in particular that : 1) a large overcharging occurs only for $T>T_{LCST}$ where bare microgels collapse and are characterized by high electrophoretic mobility; 2) charge inversion occurs at a polyelectrolyte concentration that depends on the microgel swelling and follows VPT; 3) polyelectrolyte adsorption gives rise to a reentrant condensation of microgels for $T \approx T_{LCST}$, as opposed to a continuous enhancement of particle condensation observed for monovalent salt.

\section{Experimental}
\subsection{Materials}
PNiPAM microgels are synthesized in free-surfactant emulsion-polymerization. A 1-liter three-necked round bottom flask reactor is equipped with a stirrer, a reflux condenser, and a gas inlet. In the round bottom flask we dissolved the monomer N-isopropylacrylamide (NiPAM) (from Sigma-Aldrich) (2.31 g, 20.44 mmol) and the crosslinker N,N-methylen-bis-acrylamide (BIS) (from Sigma-Aldrich) (0.04 g, 0.26 mmol) in 225 ml pure water under stirring. The initiator potassium peroxodisulfate (KPS) (from Sigma-Aldrich) (0.09 g, 0.33 mmol) is dissolved in 25 ml pure water in a separate flask. The solution containing NiPAM and BIS is bubbled with argon for $30$ min and, after heating it up to 70 $^{\circ}$C, the initiator solution is added.
After $6$ h the dispersion is cooled to room temperature and filtered through glass wool. NaN$_3$ (2 mmol) was added to prevent bacteria growth.
Due to the use of the ionic initiator KPS the microgels carry charged groups at the dangling ends of PNiPAM-chains. Since charges are preferentially oriented towards the water phase, our synthesis conditions performed at high temperature ($T=70$ $^{\circ}$C $>$ $T_{LCST}\sim$33 $^{\circ}$C), where PNiPAM is  in a globular state, effectively forces the charges to be preferentially located at the outer edge of the microgels.

$\epsilon$-Poly-L-lysine ($\epsilon$-PLL) was a kind gift from Chisso Corporation (Yokohama, Japan). This polymer, consisting of 25 to 35 L-lysine residues (Mw $\approx$ 4 kDa) is produced by a mutant of Streptomyces albulus NBRC14147 strain \cite{Hiraki00,Hamano14}, and is used as a food preservative in several countries for its antimicrobial activity against a spectrum of microorganisms, including bacteria and fungi \cite{Yoshida03}. $\epsilon$-PLL is a hydrophilic cationic homo-poly-amino acid with an isoelectric point around pH=9.0 and is described as having a peptide bond between carboxyl groups and $\epsilon$-amino groups of L-lysine residues rather than the conventional peptide bonds linking $\alpha$-poly-L-lysine ($\alpha$-PLL) \cite{hyldgaard_antimicrobial_2014} in which hydrophobic methylene side groups are fully exposed to water and may interact hydrophobically.

The diameter of the chain is approximately 0.7 nm \cite{Huang13},  the length of the monomer can be estimated as a sum of the atomic covalent radii, which gives $\approx 0.6$ nm, so that the contour length $L$ of the polymer (25-35 monomers) is $\approx$ 15-20 nm \cite{heyrovska2008}.

$\epsilon$-PLL was in the basic form and was converted to Cl salt by titration with HCl followed by extensive dialysis to eliminate the H$^+$ excess.

Hereafter, in order to quantify the charge balance in the mixtures of PNiPAM microgels and $\epsilon$-PLL, we will use the charge ratio $\xi$ defined as the nominal molar ratio $n_{lys}/n_{K^+}$, where $n_{K^+}$ is the number of moles of $K^+$ ions carried by the KPS initiator embedded in PNiPAM microgels and $n_{lys}$ the number of moles of lysine monomers dispersed in the mixtures.

\subsection{Preparation of microgel-polyion complexes}\label{preparation}
Each microgel-$\epsilon$-PLL mixture was prepared according to the following standard protocol, which was well assessed in our past investigations on liposome-polyelectrolyte complexes (see for example \cite{sennato_salt-induced_2016}).
A volume of 0.5 ml of the $\epsilon$-PLL solution at the required concentration was added to an equal volume of the microgel suspension in a single mixing step and gently agitated by hand. Before mixing, both suspension and polyelectrolyte solution were kept at room temperature to avoid interference of thermal gradients during the following measurement. After mixing the two components, the sample was immediately placed in the thermostatted cell holder of the instrument for the measurement of the electrophoretic mobility and the size of the resulting complexes.

\subsection{Viscosimetry}\label{visco}
Viscosity measurements were performed using an Anton Paar Lovis 2000 ME micro-visco\-si\-meter to obtain the constant of proportionality between PNiPAM mass fraction, $c$, and microgel volume fraction, $\varphi$, at $T=20$ $^{\circ}$C. In the range $6.25\cdot 10^{-5} < c < 7.48 \cdot 10^{-4}$ the viscosity $\eta$ of the suspensions increases linearly with $c$. Since microgels are highly swollen, their mass density is essentially the same as that of the solvent. Consequently, weight fraction $c$ and volume fraction $\varphi$ are proportional, i.e. $\varphi = kc$. We determined the constant $k$ using the $c$-dependence of the zero-shear viscosity in the dilute regime \cite{truzzolillo_bulk_2015-1}. Briefly, we determined the constant $k$ by matching the concentration dependence of the zero shear viscosity to the one predicted in the dilute regime by Einstein's formula:
\begin{equation}\label{eta}
\frac{\eta}{\eta_0}=1+2.5\varphi=1+2.5kc
\end{equation}
where $\eta_0$ is the viscosity of the solvent. By fitting $\eta/\eta_0$ to a straight line, we obtained $k = 23.9 \pm 1.3$ that allows defining the microgel volume fraction as $\varphi (T)=kc$ $R_h^3(T)$/$R_h^3(20^{\circ}$C$)$, where $R_h(T)$ is the hydrodynamic radius of the microgels measured by dynamic light scattering.

\subsection{Light scattering and electrophoretic mobility measurements}\label{size-zeta meas}
We measured the gyration radius of the bare microgels as a function of temperature by means of static light scattering. The light intensity $I(q)$ scattered by very dilute samples ($\varphi=0.001$) was measured at different scattering angles using an Amtec-goniometer. Here $q = 4\pi n \lambda^{-1} \sin(\theta/2)$ is the scattering vector, with $\lambda = 532.5$ nm the wavelength of the incident laser radiation, $n$ the solvent refractive index and $\theta$ the scattering angle. From the time averaged scattering intensity $I(q)$ the radius of gyration $R_g$ has been determined by using the Guinier approximation $I(q)=I(0)\exp[-(qR_g)^2/3]$ \cite{guinier_small_1955}.

The hydrodynamic size and the size distribution of microgels and polyion-microgels complexes were characterized by means of dynamic light scattering measurements (DLS), employing a MALVERN Nano Zetasizer apparatus equipped with a 5 mW HeNe laser (Malvern Instruments LTD, UK). This system uses backscatter detection, i.e. the scattered light is collected at an angle of 173$^{\circ}$. The main advantage of this detection geometry, when compared to the more conventional 90$^{\circ}$, is its inherent larger insensitiveness to multiple scattering effects \cite{dhadwal_fiberoptic_1991}. Intuitively, since nor the illuminating laser beam, nor the detected scattered light need to travel through the entire sample, chances that incident and scattered photons will encounter more than one particle are reduced. Moreover, as large particles scatter mainly in the forward direction, the effects on the size distribution of dust or, as is our case, of large irregular aggregates (lumps or clots),  are greatly reduced. To obtain the size distribution, the measured autocorrelation functions were analyzed by means of the CONTIN algorithm \cite{provencher_constrained_1982}. Decay times are used to determine the distribution of the diffusion coefficients $D_0$ of the particles, which in turn can be converted in a distribution of apparent hydrodynamic diameter, $D_h$, using the Stokes-Einstein relationship $D_{h} = k_BT/3\pi\eta D_0$, where $k_B$ is the Boltzmann constant, $T$ the absolute temperature and $\eta$ the solvent viscosity.\\
The values of the radii shown in this work correspond to the average values on several measurement and are obtained from intensity weighted distributions \cite{provencher_constrained_1982,de_vos_quantitative_1996}.

The electrophoretic mobility of the suspended microgels was measured by means of the same NanoZetaSizer apparatus employed for DLS measurements. This instrument is integrated with a laser Doppler electrophoresis technique, and the particle size and electrophoretic mobility can be measured almost simultaneously and in the same cuvette. In this way, possible experimental uncertainties due to different sample preparations, thermal gradients and convection are significantly reduced. Electrophoretic mobility is determined using the Phase Analysis Light Scattering (PALS) technique \cite{tscharnuter_mobility_2001}, a method which is especially useful at high ionic strengths, where mobilities are usually low. In these cases the PALS configuration has been shown to be able to measure mobilities two orders of magnitudes lower than traditional light scattering methods based on the shifted frequency spectrum (spectral analysis).
All DLS and electrophoretic measurements were performed at fixed microgel concentration $c=0.001$ wt/wt (i.e. $\varphi(20^{\circ}$C$)=0.024$ at 20 $^{\circ}$C).
The used thermal protocol consists of  an ascending ramp from 20 $^{\circ}$C to 40 $^{\circ}$C with temperature step of 1 $^{\circ}$C. At each  step, samples have been left to thermalize 300 s at the target temperature, then measurement of electrophoretic mobility  and size have been performed.

\subsection{Transmission electron microscopy}
Transmission electron microscopy (TEM) was used to study the morphology of PNiPAM and PNiPAM-PLL complexes. All the samples for TEM measurements have been prepared by depositing 20 $\mu$l of microgel suspensions ($\varphi=0.024$) on a 300-mesh copper grid for electron microscopy covered by a thin amorphous carbon film. Samples have been deposited both at room temperature and at 40$^\circ$C in order to reveal morphological differences induced by temperature. To prepare PNiPAM samples above the VPT, both PNiPAM suspension,  TEM grids and pipette tips have been heated at 40$^\circ$C. For TEM observation of PNiPAM-PLL complexes samples were prepared at the same concentration of the samples investigated by DLS and electrophoretic mobility. After withdrawal of 20 $\mu$l aliquot of the mixed PNiPAM-PLL suspension, the thermal protocol from 20$^\circ$C to 40$^\circ$C was used to promote the formation of complexes.
At 40 $^{\circ}$C, an aliquot of this PNiPAM-PLL sample was withdrawn and deposited on the pre-heated TEM grid in a thermostatted oven. After 5 minutes drying in the oven, the samples were dried by filter paper. When necessary, negative staining was realized by addition of 10 $\mu$l of 2 $\%$ aqueous phosphotungstic acid (PTA) solution (pH-adjusted to 7.3 using 1 N NaOH). Measurements were carried out by using a FEI TECNAI 12 G2 Twin (FEI Company, Hillsboro, OR, USA), operating at 120 kV and equipped with an electron energy loss filter (Biofilter, Gatan Inc, Pleasanton, CA, USA) and a slow-scan charge-coupled device camera (794 IF, Gatan Inc, Pleasanton, CA, USA).

\subsection{Dielectric Spectroscopy}\label{dielectric}
Dielectric spectroscopy (DS) experiments were performed using three different setups probing three partially overlapping frequency ranges. In all cases, the temperature of the cells was controlled through a Haake K35/D50 circulating water bath, which allows for a temperature control within 0.1 $^\circ$C.

In the low (40 Hz $\leq\nu\leq$ 100 MHz) and intermediate (1 MHz $\leq\nu\leq$ 1.8 GHz) frequency ranges, measurements were performed through impedance analyzers (Hewlett-Packard, model 4294A and model 4291A, respectively). In these cases the dielectric cells consist of a short section of a cylindrical coaxial cable (inner radius 1.5 mm, outer radius 3.5 mm) connected to the meter by means of a precision APC7 connector. Further details are given in refs. \cite{bordi_01,bordi_04}.\\
At the higher frequencies (40 MHz to 40 GHz) we employed a homemade cell for liquid samples connected to a vector network analyzer (VNA, Anritsu 37297D) through a microwave line. The dielectric cell is build up with a gold plated brass cylinder of inner radius 1.5 mm, 10 mm long, with a commercial glass bead transition (Anritsu K100) that closes its lower end.
The chosen value of the inner radius is the result of a balance between the request of a high enough cutoff frequency and the requirement of a sufficiently large cavity to avoid the retention of bubbles when the cell is filled, especially in the case of rather viscous microgel suspensions.

The most relevant data are limited to the MHz range (1 MHz $\leq\nu\leq$ 1 GHz), where the relaxations due to the microgels and the polymer were detected. However, the availability of a wide enough low frequency tail allows for a far better correction of the electrode polarization effects, while the high frequency tail, dominated by solvent contribution, allows for a more accurate definition of the spectrum dominated by the microgels and the polyelectrolyte.

The electrode polarization contribution has been subtracted following the procedure described in Bordi et al.\cite{bordi_01}. In particular, we assume that this contribution can be represented as an impedance $Z=K^{-1} (i\omega)^{-\alpha}$ of the cell (Constant Phase Angle (CPA) approximation). In order to determine the parameters $K$ and $\alpha$ the low tails of the spectra are fitted by assuming that the complex permittivity of the solution can be written as $\varepsilon=\varepsilon_s+i\sigma_{dc}/(\varepsilon_0\omega)$ in the region where the divergence of the real part of the measured permittivity is systematically observed ($\nu< 10^5$ Hz). Once the two parameters ($K$ and $\alpha$) are obtained, the electrode polarization contribution is fully determined and can be algebraically subtracted from the overall curve. In this way, we also determine the dc conductivity  $\sigma_{dc}$, which we compare and systematically find in reasonable agreement with the conductivity measured through the low-frequency potentiometer embedded in the Nano Z-sizer used for the electrophoretic characterization of the samples.


The spectra have been analyzed as follows: 1) the electrode polarization contribution, determined by fitting the low frequency tail with the CPA expression, has been properly subtracted from the measured curve; 2) the corrected spectra have then been analyzed in terms of  the Looyenga equation \cite{Looyenga_65,bordi_02}

\begin{equation}\label{Looyenga_eq}
\varepsilon(\omega)^{1/3}=\tilde\epsilon_p(\omega)^{1/3}\varphi+(1-\varphi)\tilde\epsilon_m (\omega)^{1/3}
\end{equation}
where $\varepsilon$ is the total permittivity of the solution, $\tilde\epsilon_p$ is the effective permittivity of the colloid, $\tilde\epsilon_m$ is the permittivity of the solvent, $\varphi$ is the volume fraction occupied by the colloids, and $\omega$ is the radian frequency of the imposed electric field. All the permittivities in the above expression are complex quantities.
Having determined the solvent permittivity, from equation \ref{Looyenga_eq} the effective permittivity of the colloid can be obtained once $\varphi$  has been measured by viscosimetry as described in section \ref{visco}.
The nominal charge ratio was tuned by varying microgel concentration in the range $0.14 \leq\varphi$(20$^{\circ}$C)$\leq 0.56$ at one $\epsilon$-PLL concentration (4.4 mg/ml).

\section{Results and discussion}
\subsection{Characterization of bare microgels}\label{barepnipam}
\begin{figure}[htbp]
  \includegraphics[width=7.5cm]{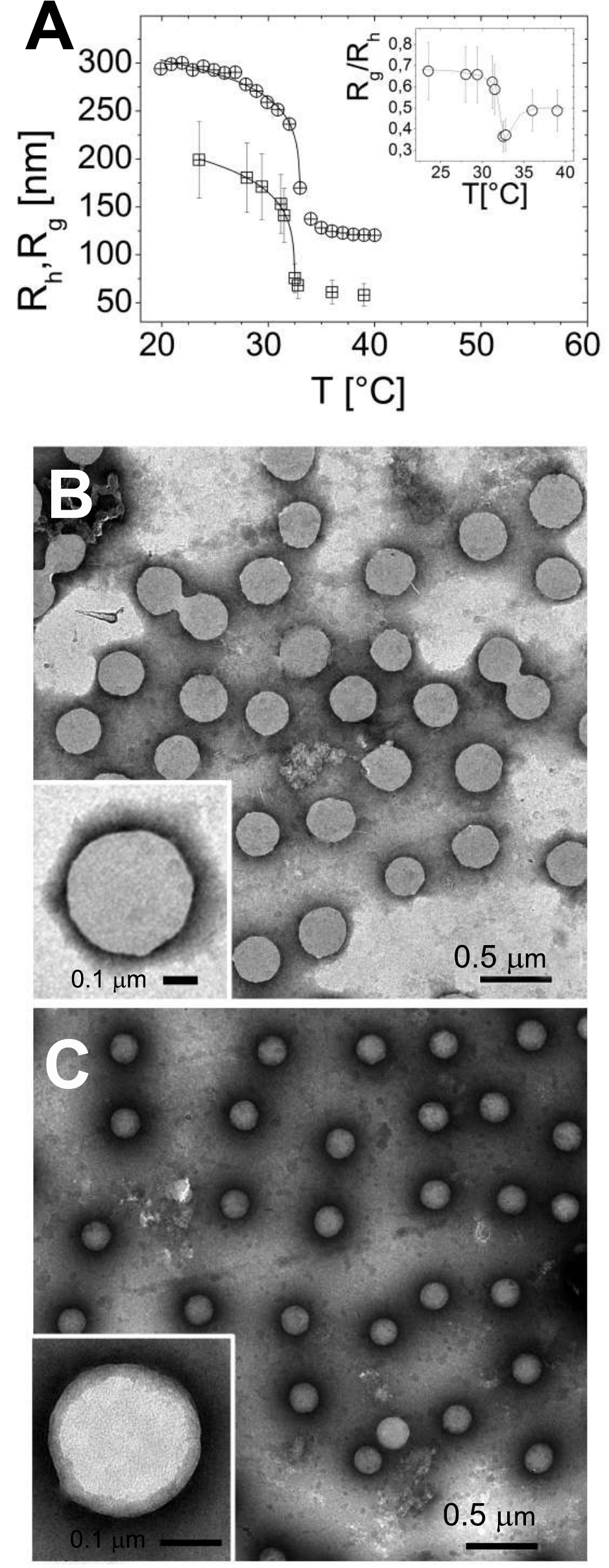}\\
  \caption{Hydrodynamic $R_h$ (circles) and gyration radius $R_g$ (squares) of PNiPAM microgels as a function of temperature obtained by DLS and SLS respectively (panel A). The inset of panel A shows the ratio $R_g/R_h$ as a function of temperature. TEM images obtained by negative PTA staining of PNiPAM microgels prepared at room temperature (panel B) and heated at 40 $^{\circ}$C (panel C). A significant reduction of the microgel size is observed.  TEM images of single microgels prepared at room temperature and heated at  40 $^{\circ}$C are shown in the insets of panels B and C, respectively.}\label{pnipam-ch SIZE}
\end{figure}

The particle size, as measured both by $R_h$ and $R_g$, shrinks as the temperature is raised above the LCST (Figure \ref{pnipam-ch SIZE}-A), and below LCST both $R_h$ and $R_g$  are well fitted by a critical-like function \cite{bischofberger_new_2015} $R_{h,g}=R_0(1-T/T_c)^{\alpha}$. We obtained $R_0=331 \pm 4$ nm, $T_c=33.03 \pm 0.04$ $^\circ$C , and $\alpha=0.096 \pm 0.008$ for $R_h$ and $R_0=241 \pm 72$ nm, $T_c=32.51 \pm 0.02$ $^{\circ}$C and $\alpha=0.14 \pm 0.01$ for $R_g$.
In agreement with previously reported results the two radii differ significantly. This difference has been attributed to an uneven distribution of crosslinks within the microgel, giving rise to a core-shell structure \cite{bischofberger_new_2015,senff_temperature_1999,mason_density_2005,reufer_temperature-sensitive_2009}. In particular, for all temperatures we find $R_g/R_h<0.77$, that is the value expected for homogenous spheres, which points out that the distribution of monomer density is peaked at the center of the microgel \cite{bischofberger_new_2015, sessoms_multiple_2009, Vogel2017}. Moreover, temperature affects differently the core and the periphery of the microgels as signaled by the minimum of $R_g/R_h$ for $T \sim T_c$ (inset of figure \ref{pnipam-ch SIZE}-A), suggesting that close to the critical temperature the disuniformity of the microgels is maximum. Such finding, already observed for other thermosensitive microgels \cite{arleth_volume_2005,sun_investigation_2005}, suggests that critical fluctuations at $T \sim T_c$ favors the shrinkage of the core, leaving far apart the dangling ends bearing the majority of the microgel charges.\\
The change of size of microgels can be also inferred by TEM images. In figure \ref{pnipam-ch SIZE}, we show the microgels prepared at room temperature ($T<T_c$) and at 40 $^{\circ}$C stained with PTA (negative contrast) (panels B and C, respectively). The reduction of size due thermal transition can be clearly appreciated by comparing the two images.  With the negative contrast technique the microgel particles appear as light grey objects since they are impenetrable to PTA.

\begin{figure}[htbp]
  \includegraphics[width=10cm]{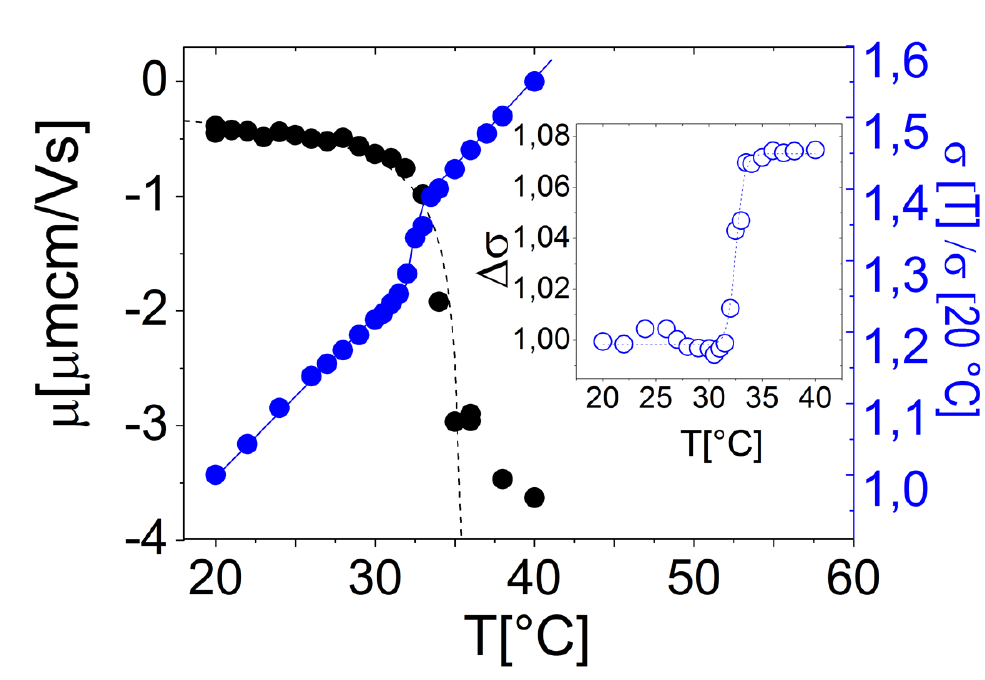}\\
  \caption{Electrophoretic mobility $\mu$ (left axis) and typical behavior of the low-frequency limit conductivity of a suspensions of PNiPAM microgels (right axis) as a function of temperature. $\mu$ was measured at the same concentration as in DLS measurements ($\varphi=0.024$).
  Conductivity values are normalized to the value measured at T=20 $^\circ$C, and are measured at $c=5$ w/w ($\varphi=1.19$), however, in the whole range of  investigated concentrations (from $\varphi=1.19$ down to $\varphi=0.024$) the behavior is qualitatively similar. The inset shows the conductivity jump $\Delta\sigma$ after the subtraction of the linear trend $\sigma_l(T)$ (see text).}\label{pnipam-ch CHARGE}
\end{figure}

In the insets two single microgel particles, prepared respectively at room temperature and at 40 $^{\circ}$C, are shown. The heterogenous structure of our microgels is also apparent in TEM images (inset of figure \ref{pnipam-ch SIZE}-C): PTA, which accumulates close to microgel forming an external black halo at the microgel interface, apparently penetrates a small distance inside the crosslinked network giving rise to an almost regular dark gray corona,  about $20-25$ nm thick, pointing out the presence of a less tangled shell in the periphery of the microgel. Conversely, in our images, an homogeneous appearance is found for microgels prepared at room temperature (inset of figure \ref{pnipam-ch SIZE}-B).
Figure \ref{pnipam-ch CHARGE} shows the typical behavior of electrophoretic mobility $\mu(T)$ and  electrical conductivity of microgel suspensions as a function of the temperature. Electrophoretic mobility is measured at $c=0.1$ wt/wt, which is also the concentration used for the size and electrophoretic characterization of the microgels in the presence of added salt and $\epsilon$-PLL chains. As expected, $\mu(T)$ is affected by the VPT and decreases unambiguously as the temperature is increased above $T>T_c$. However, the vertical drop of the mobility is observed slightly  above the critical  temperature $T_c\simeq 33$ $^{\circ}$C. In fact, at $T=T_c$ the mobility is only $\approx$2 times lower than at 20 $^{\circ}$C. As for the microgel radii, we quantify the mobility drop up to an 'electrokinetic transition temperature' $T_{c\mu}$ by using a critical-like function $\mu(T)=\mu_0(1-T/T_{c\mu})^{-\alpha}$, obtaining from the fit the parameters $\mu_0=-0.22 \pm 0.03$ $\mu$m cm/Vs, $T_{c\mu}=35.7 \pm 0.4$ $^{\circ}$C and $\alpha=0.59 \pm 0.4$. It is worth noting that the difference between the critical temperature associated to the VPT and that associated to the electrokinetic transition is $\Delta=T_{c\mu}-T_c \gtrsim 2.7$ $^{\circ}$C. Such significant difference between the two transition temperatures has been already discussed by Pelton et al. \cite{pelton_particle_1989} and by Daly et al. \cite{daly_temperaturedependent_2000} and has been attributed to a multi-step transition, where the almost-uncharged core collapses first, with a significant reduction of particle size, and the shell, where the charges are mostly confined, collapsing only at higher temperature. This picture is fully consistent with the minimum of $R_g/R_h$ that we observe at $T \sim T_c$.
The VPT of microgels has also a detectable effect on the low frequency limit of the electrical conductivity $\sigma(T)$ of the suspensions. Figure \ref{pnipam-ch CHARGE} shows the typical behavior of $\sigma(T)/\sigma($20$^{\circ}$C$)$ vs T. In this example the microgel concentration is quite high, $c=5$ w/w ($\varphi=1.19$), but in the whole range of investigated concentrations (from $\varphi=1.19$ down to $\varphi=0.024$) the behavior is qualitatively similar. In all cases, superimposed to the linear trend $\sigma_l(T)=\alpha+\beta T$ due to the electrolyte contribution predicted by the Fuoss-Onsager theory \cite{AqueousElectrolyteSolutions07}, there is a sudden increase of the conductivity at the VPT (inset of figure \ref{pnipam-ch CHARGE}), only the magnitude of the jump $\Delta\sigma=\sigma(T)/\sigma_l(T)$ being dependent on the microgel concentration.
Such a sharp increase of $\sigma(T)$ can be explained in terms of the simultaneous sharp decrease of the suspension viscosity due to  the reduction of the microgel volume fraction, and/or attributed to an increase of the microgel charge density driven by the VPT, and the partial expulsion of condensed counterions from the inner part of the microgels, with a consequent increase of their effective charge above the VPT. However the latter hypothesis, conforming to a reduced counterion condensation on the microgels, is in contrast with recently published results\cite{braibanti_impact_2016} suggesting that the effective charge of PNiPAM microgels is an increasing function of their size. Therefore, both the mobility and the conductivity increase seem rather the result of particle shrinkage, that causes a net increase of particle charge density, and a large concomitant increase of free space. In fact it's worth noting that the reduction of the particles radius of a factor 2-2.5 above the VPT (Figure \ref{pnipam-ch SIZE}-A) implies a corresponding reduction of a factor $\approx 10$ of the volume fraction $\varphi$, so that e.g. in the case of the sample shown in figure \ref{pnipam-ch CHARGE}, the free space changes from virtually zero below the VPT ($\varphi \gtrsim 1$), where the suspension is completely  jammed, to $\approx 90 \%$ above the VPT.

The next sections will be devoted to the description of the general phenomenology stemming from the addition of a uni-univalent inorganic electrolyte, NaCl, and a cationic polyelectrolyte ($\epsilon$-PLL) in diluted microgel suspensions.

\subsection{Effect of monovalent salt}\label{salt}

\begin{figure}[htbp]
  \includegraphics[width=10cm]{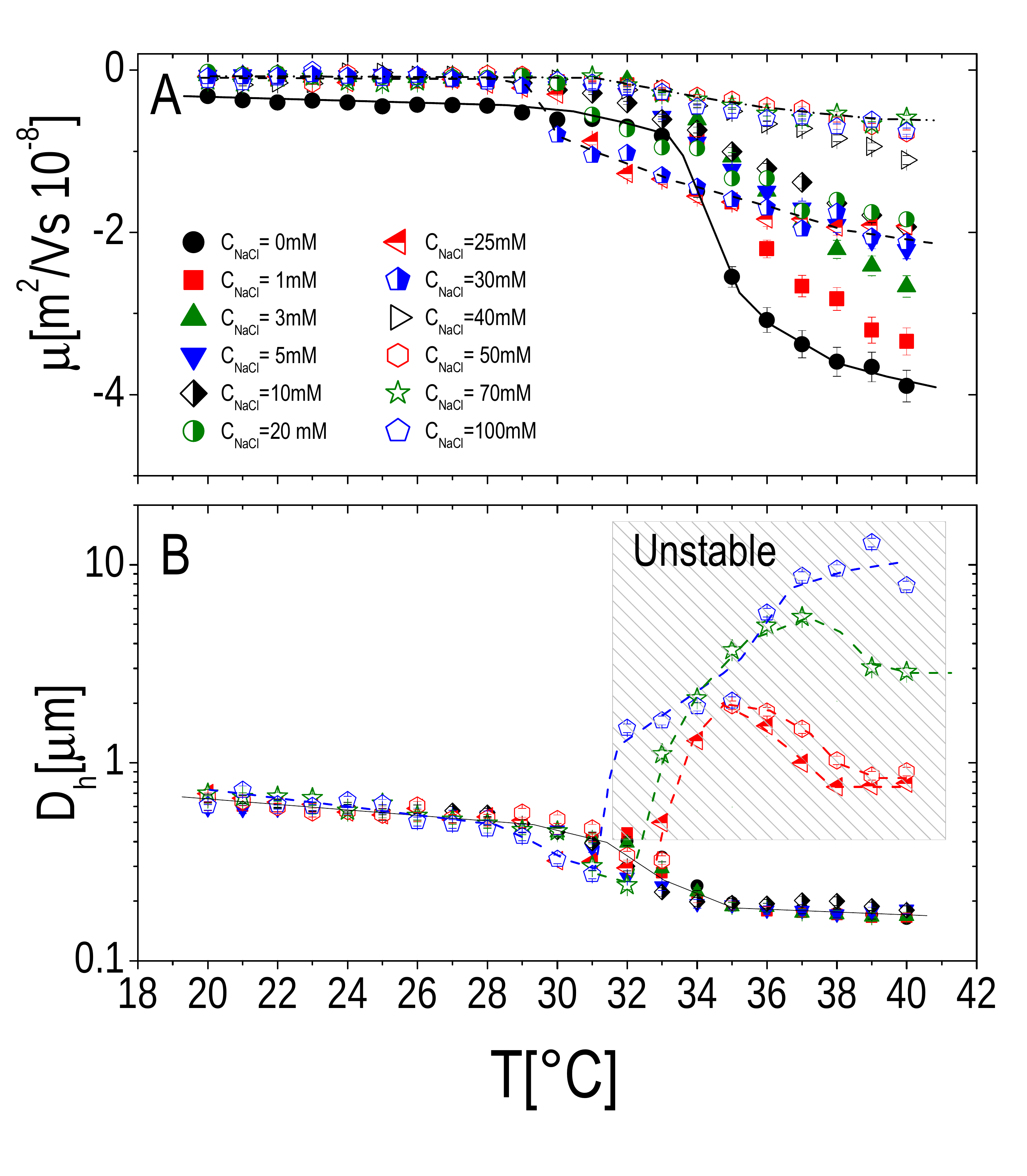}\\
  \caption{Electrophoretic mobility $\mu$ (panel A) and hydrodynamic diameter $D$ (panel B) of PNiPAM microgels as a function of temperature for different salt concentrations as indicated in the figure. The shaded region in panel B denotes all the samples where flocculation has been observed after 12 hours (Empty points in both panels). In panel A lines are drown to guide the eye though three selected data sets: $C_{NaCl}=0$ mM, $C_{NaCl}=30$ mM and $C_{NaCl}=70$ mM.}\label{NaCl-Panel}
\end{figure}

The effect of monovalent salt (NaCl) has been investigated by monitoring the electrophoretic mobility and the hydrodynamic diameter as a function of temperature and by varying the salt concentration $C_{NaCl}$. Measurements were made by progressively heating the sample in the presence of a constant salt concentration ranging from $0$ mM to $50$ mM (Figure \ref{NaCl-Panel}), at a fixed microgel concentration c = 0.001 wt/wt.
All the mobility curves exhibit the same trend: at low temperatures the electrophoretic mobility remains unaffected by temperature at any salt concentration, while  for temperatures higher than $T_{c\mu}$ it decreases (in absolute value) down to values depending on $C_{NaCl}$.
As for bare microgels in the absence of added salt $|\mu|$ rises with temperature owing to an increase of the surface-charge density, whose effect dominates over the enhanced friction forces at work when particles shrink and their monomer density increases.

By monitoring the hydrodynamic diameters vs temperature, we observe the formation of aggregates above VPT for $C_{NaCl} \geq$ 25  mM (see figure \ref{NaCl-Panel}-B): in this range of salt concentrations, an increase of temperature above $T_c$ triggers the formation of aggregates, whose size decreases as temperature is further raised up to 40 $^{\circ}$C, due to the single particle shrinkage. These aggregates however are not stabile, since a distinct  flocculation is observed for $C_{NaCl}>$25 mM and $T\gtrsim T_c$ after approximately 12 hours. This region of the $C_{NaCl}$-$T$ plane must be then considered unstable.
The absence of flocculation for $C_{NaCl}<25$ mM is in agreement with the flocculation behavior of similar PNiPAM microgels observed by Rasmusson et al.\cite{rasmusson_flocculation_2004} who pointed out that the charge carried by the microgels (due to KPS initiator in our case) is sufficient for their stability in the temperature range $20$ $^{\circ}$C$\leq T\leq$60 $^{\circ}$C for $C_{NaCl}<$25 mM.
Microgel aggregation caused by the reduced solvent quality above the VPT and salt addition has been widely discussed in some previous works\cite{liao_fractal_2012,shen_colloidal_2012,lopez-leon_thermally_2014,rasmusson_flocculation_2004} and will not be further examined here.\\

\begin{figure}[htbp]
  \includegraphics[width=10cm]{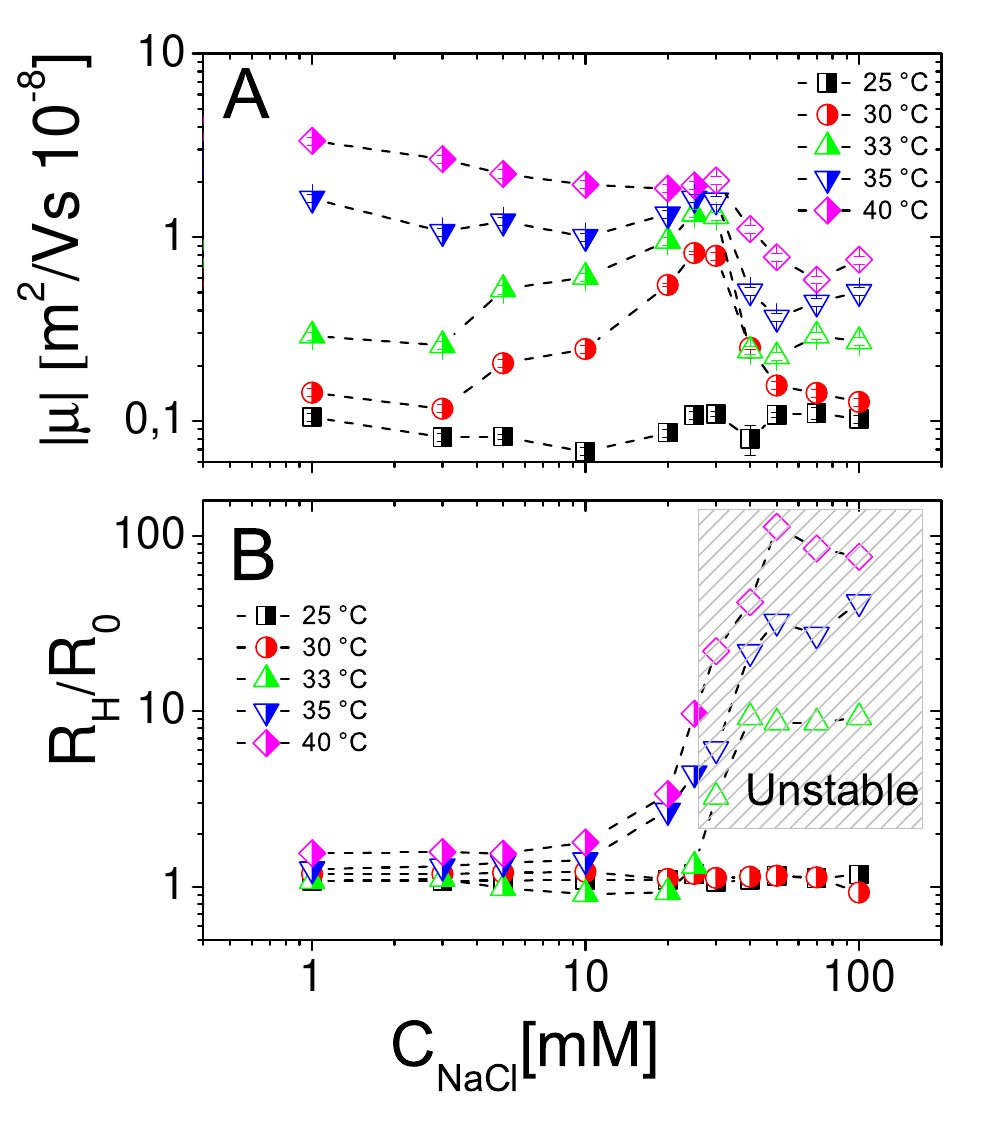}\\
  \caption{Electrophoretic mobility modulus $|\mu|$ (panel A) and normalized hydrodynamic radius $R_h/R_0$ (panel B) of PNiPAM microgels as a function of salt concentration $C_{NaCl}$ for selected temperatures as indicated in the figure. The shaded region in panel B encloses all the samples (empty points) where flocculation has been observed after 12 hours.}\label{NaCl-Panel-2}
\end{figure}

On the contrary it is worth to point out the non-trivial dependence of the electrophoretic mobility on $C_{NaCl}$ at different temperatures (Figure \ref{NaCl-Panel-2}-A) since, to our knowledge, this is an aspect that has not been previously discussed. Actually, well below the VPT the microgel mobility is nearly unaffected by the addition of monovalent salt. Indeed, in the case of swollen microgels the particle/solvent interface is poorly defined and characterized by a low charge density. Here the classical electric double layer description, that leads to the Smoluchowski equation predicting a scaling $|\mu| \sim C_{NaCl}^{-1/2}$ \cite{hunter_zeta_1981}, cannot be applied. A weaker than expected dependence has been already found by Sierra-Martin et al.\cite{sierra-martin_thermal_2006} who reported $|\mu|\sim C_{NaCl}^{-0.34}$ for similar microgels below VPT.
As the temperature is raised, the particle surface becomes better defined and the overall $|\mu|$ behavior resembles that of compact hard particles, being characterized by a pronounced maximum at $C_{NaCl} \approx 30 $ mM. The presence of a maximum is predicted by standard electrokinetic models taking into account retardation forces due to double layer relaxation\cite{obrien_electrophoretic_1978,zhou_computer_2015} around hard spheres for sufficiently short screening lengths ($R_h/\lambda>3$, where $\lambda$ is the Debye screening length). There are essentially four forces accounted for in these models that determine the steady velocity of a particle subject to an external electric field: 1) the electric force acting on the colloid; 2) a hydrodynamic drag force; 3) a further electrostatic contribution due to the ion cloud displacement with respect to the center of the colloid; 4) a relaxation force, hydrodynamic in origin, resulting from the ion motion altering the solvent flow velocity around the particle. The non-monotonic behavior of the mobility can be understood by considering the competition between the electric and the relaxation force. Indeed the former scales as $\lambda\sim C_{NaCl}^{-1/2}$, while the latter scale as $\lambda^2$ \cite{obrien_electrophoretic_1978,zhou_computer_2015,sierra-martin_thermal_2006}. With decreasing salt concentration $|\mu|$ increases until the faster growing relaxation forces take over, determining the decrease of the mobility. This is the case of our microgels at $T\geq30$ $^{\circ}$C, where mobility is non-monotonic and for which $44 \lesssim R_h/\lambda \lesssim 123$.
Also, the reduced extent of the mobility maximum observed at $T>33$ $^{\circ}$C, where microgel collapse occurs, conforms to an enhancement of the electric force whose screening dominates over the suppression of the relaxation drag force and determines a continuous decrease of the mobility.

In addition to that, the existent theories\cite{ohshima_electrophoresis_1995,Fujita57,hermans_sedimentation_1955,Ohshima02} considering electrophoretic retardation forces predict that the mobility of soft penetrable particles does not depend on the size (or aggregation number), being uniquely determined by their charge density and electrophoretic friction. The latter are two intensive quantities that stay constant during any ongoing aggregation process at fixed temperature. For this reason we may expect negligible effects of aggregation on the measured mobility. However the relaxation drag force arising from the ion flow through the microgels is not taken into consideration by such theories and may give a size-dependent contribution to the total drag force. As we show later (section \ref{revtherm}) this is ruled out for our microgels.

We may also wonder if the increase of $C_{NaCl}$ reduces the solvent quality and affects the microgel charge distribution. In figure \ref{NaCl-Panel-2}-B we show the hydrodynamic radius $R_h$ of microgels as a function of $C_{NaCl}$ normalized to the radius $R_0$ measured in salt-free water for different temperatures. In our case salt addition does not significantly affect the size of the single microgels before clustering occurs, for this reason an effect of charge density enhancement due to a weak particle deswelling is ruled out. 

Therefore, in agreement with previous works, the electrophoretic behavior of our PNiPAM microgels spans from that of soft, swollen and weakly charged particles to that predicted for hard charged colloids, and represents in this work an important frame of reference for our investigation of the effect of small polyions on the stability and the dielectrophoretic behavior of these thermoresponsive colloids.
\subsection{Polyelectrolyte-microgel complexation}\label{microPLL}
\begin{figure}[htbp]
  \includegraphics[width=12cm]{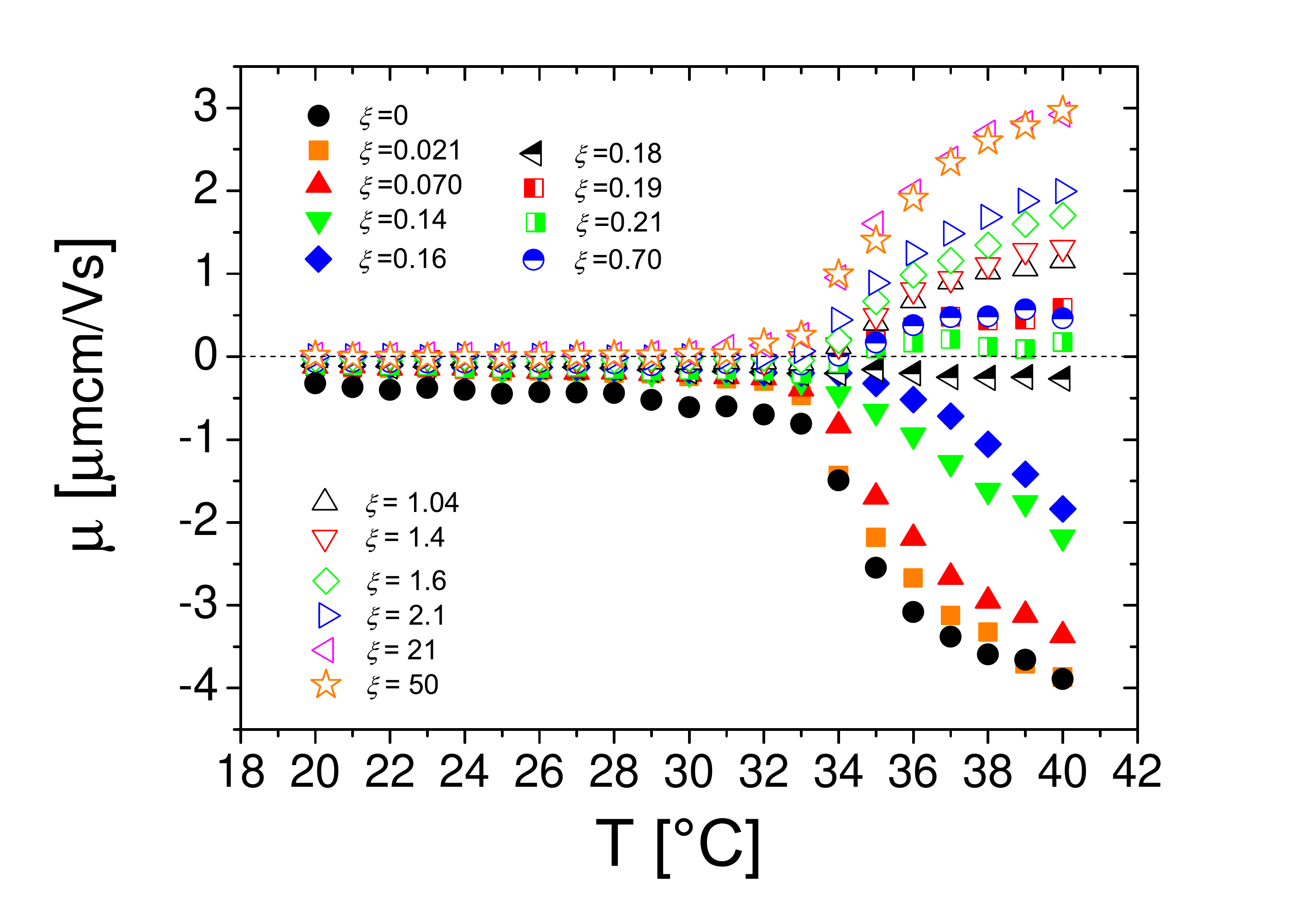}\\
  \caption{Electrophoretic mobility $\mu$ of PLL-microgel complexes as a function of temperature for different PLL concentrations (charge ratios $\xi$) as indicated in the figure.}\label{mobility-panel}
\end{figure}

\begin{figure}[htbp]
  \includegraphics[width=12cm]{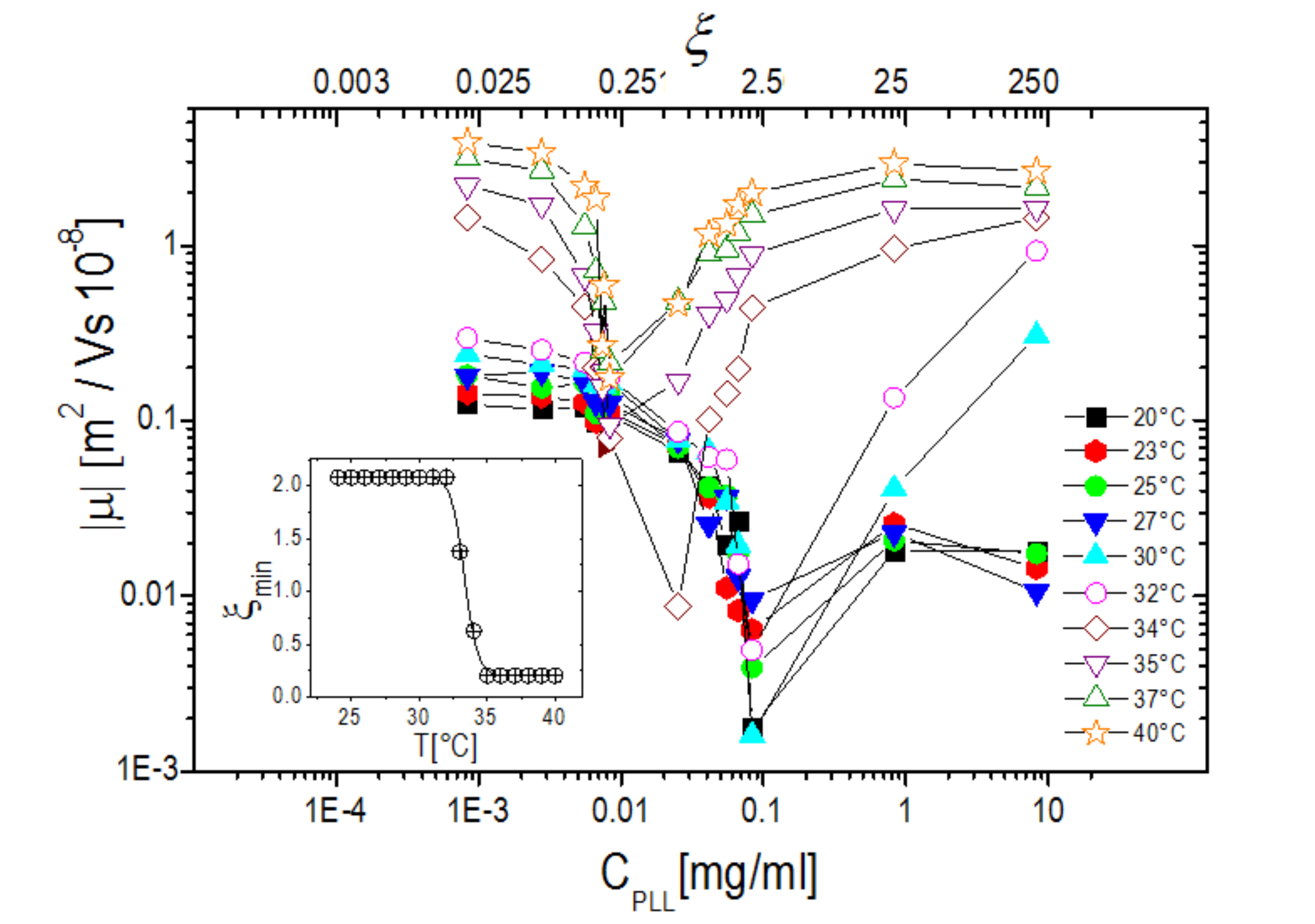}\\
  \caption{The absolute value of electrophoretic mobility $|\mu|$ of PLL-microgel complexes as a function $C_{PLL}$ (charge ratio $\xi$) for selected temperatures as indicated in the figure. The inset shows the charge ratio $\xi_{min}$ where the mobility modulus reaches its minimum value.}\label{mobility-CPLL}
\end{figure}

The behavior of the electrophoretic mobility $\mu$ of PNiPAM-PLL complexes is shown in figure \ref{mobility-panel}. This behavior is not exactly  what one would expect, since according to the classical Gouy-Chapman-Stern theory \cite{israelachvili_intermolecular_1985}, the Ohshima's \cite{ohshima_electrophoresis_1995, Ohshima02} and Hermans-Fujita's \cite{Fujita57,hermans_sedimentation_1955} equations, the latter more specifically valid for penetrable particles, the absolute value of $\mu$ should decrease as the ionic strength of the solution increases.
Analogously to what has been observed for microgels in the presence of monovalent salt, we may distinguish two regimes delimited by the microgel electrokinetic transition (ET). For $T<T_{c\mu}$ $\mu$ is low and depends very weakly on temperature and $\epsilon$-PLL concentration. On the contrary, for $T>T_{c\mu}$ the mobility is dramatically affected by both temperature and polyelectrolyte concentration: as the $\epsilon$-PLL content is increased $\mu$ passes from largely negative values ($\mu(40 ^{\circ}C)=-3.63$ $\mu$mcm/Vs for $\xi=0$) to largely positive values ($\mu(40 ^{\circ}C)=2.919$ $\mu$mcm/Vs for $\xi=21$).\\
This temperature-dependent overcharging of PNiPAM-PLL complexes points out the importance of the microgel VPT for the adsorption of PLL chains, suggesting that the net charge of the polyelectrolyte-microgels complexes can be finely adjusted by both changing temperature and polyelectrolyte content.\\
Figure \ref{mobility-CPLL},  where the modulus of the mobility $|\mu|$ is plotted versus the polymer concentration for different temperatures, shows more clearly the neutralization and the overcharging of the complexes.
The existence of an isoelectric point is marked by the minimum of $|\mu|$, whose position as a function of $C_{PLL}$ (or $\xi$) allows to track the amount of $\epsilon$-PLL needed to neutralize the microgel charge. We note that: 1) also below the ET at relatively large charge ratios the overcharging of the microgels appears clearly, suggesting that the charge density of microgels in the swollen state is sufficient to promote a significant adsorption of $\epsilon$-PLL chains; 2) the  isoelectric point crossing (and overcharging) occurs for $\epsilon$-PLL concentrations that depend on temperature (inset of figure \ref{mobility-CPLL}). Indeed it is worth noting that the isoelectric point ($\xi\simeq2.1$ for $T<T_{c\mu}$) drops sharply to $\xi\simeq0.21$ as $T$ crosses the ET: less polymer is needed to neutralize the microgels when they are more densely charged.\\

Let us now discuss more in detail the overcharging of microgels following the experimental protocol described in section \ref{preparation}.
Microgels and $\epsilon$-PLL chains are mixed at T=20 $^{\circ}$C, well below VPT and ET. In these conditions, by assuming a homogeneous distribution of crosslinker, we can calculate the average microgel mesh size as $\simeq7$ nm, a value larger than both the estimated size of the PLL chains calculated in gaussian chain approximation, which is $2R_{g}\simeq 7.2$ nm, and the true mesh size of the microgel outer shell, characterized by a less dense monomer density than the core. Here $R_g$ is the gyration radius of the chains that has been estimated as $R_g=2l_p(N_k/6)^{1/2}$, where $l_p$=1.8 nm is the known persistence length measured for $\alpha$-polylysine chains \cite{brant_configuration_1965} and $N_k=6$ is the number of statistical segments.
Thus we expect that $\epsilon$-PLL chains interpenetrate inside the swollen microgel, staying confined in its periphery, where the oppositely charged sulfate groups are located.
The charge-to-charge distance on the microgels is the highest possible for $T\ll T_{\mu c}$ and hence one PE chain gets electrostatically bound to only one or a few ionized groups on the microgel, resulting in a low adsorption energy (1 $k_BT$ per ion pair).

For this reason a large number of chains, which are partially free, is needed to neutralize the microgels. By contrast for $T>T_{\mu c}$ sulfate groups are much closer to each other, the adsorption energy and number of condensed counterions increase consequently and more than one sulfate group can possibly be neutralized by one single PLL chain, pushing the isoelectric point towards lower PLL concentrations.

To corroborate such hypothesis we can give approximate upper bound values for the charge-to-charge distance within the microgel supposing that the sulfate groups are distributed within all the microgel volume. From the synthesis we know that each microgel bears $Z=3.75 \cdot 10^5$ sulfate groups and this gives an average charge-to-charge distance $d_{cc}=(D_h^3/6Z)^{1/3}$ of 6.7 nm for $T<T_{\mu c}$ and 2.7 nm for $T>T_{\mu c}$. Such distances must be compared with the size of a $\epsilon$-PLL chain ($2R_g=$7.2 nm). This calculation, although approximate, shows that the charge density variation induced by the VPT may bring a single PLL-chain to neutralize more than one sulfate group anchored to the NiPAM network and significantly reduce the amount of chains needed to neutralize the whole microgel.

Moreover it has been shown \cite{popa_thin_2007,cahill_adsorption_2008,maroni_studying_2014} that counterions provide additional screening of the electrostatic interactions between the polyelectrolytes in the lateral direction. The interaction remains approximately of the screened Coulomb type, but the effective screening length is reduced through the additional counterions within the diffuse layer. Therefore larger adsorption energies and screened lateral repulsions in the proximity of the microgel ideal surface cooperatively determine a larger fraction of adsorbed chains and cause the observed shift of the isoelectric point towards a lower value of the nominal charge ratio as temperature is raised above ET.


This scenario also conforms to the change of the adsorbent power of charged colloids predicted by scaling theories \cite{andelman_polyelectrolyte_2000}, suggesting that surface excess is ruled by surface charge density.

It is likewise worth noting that cationic  $\epsilon$-PLL chains keep on adsorbing well beyond the isoelectric point. This is indeed not surprising and systematically occurs in polyelectrolyte-colloid mixtures until a saturation threshold, specific of each system, is reached.

Far away from the isoelectric point, charge fractionalization \cite{nguyen_model_2002,zhang_phase_2005} and counterion release \cite{andelman_polyelectrolyte_2000-1,yigit_interaction_2016} mainly determine the net electrostatic attraction between PEs and microgels.

The first mechanism is very well explained in the seminal work of Nguyen and Shklovskii \cite{nguyen_model_2002}. By forming dangling ends at the particle surface, the adsorbed chains gain some conformational entropy. The charge vacancies left by these defects can be locally large enough to drive the oncoming polyelectrolyte chain nearer to the surface where, due to the repulsion between the like-charged chains, vacancies can join and enlarge, also allowing the newcomer chain to adsorb. This mechanism is likely to be present above VPTT when the mutual distance between the charges of the microgels is reduced.

A second mechanism driving the overcharging of microgels is counterion release due to PE adsorption. For $\epsilon$-PLL chains the fraction of condensed counterions according to the Manning theory is $1-b/l_b \approx 0.15$, where $b$=0.6 nm is the monomer size and $l_b$=0.7 nm is the Bjerrum length in water at $T=$ 20 $^{\circ}$C. The release of these counterions promotes polyelectrolyte adsorption far from the isoelectric point on charge-inverted microgels \cite{andelman_polyelectrolyte_2000-1,yigit_interaction_2016}.

\begin{figure}[htbp]
  \includegraphics[width=10cm]{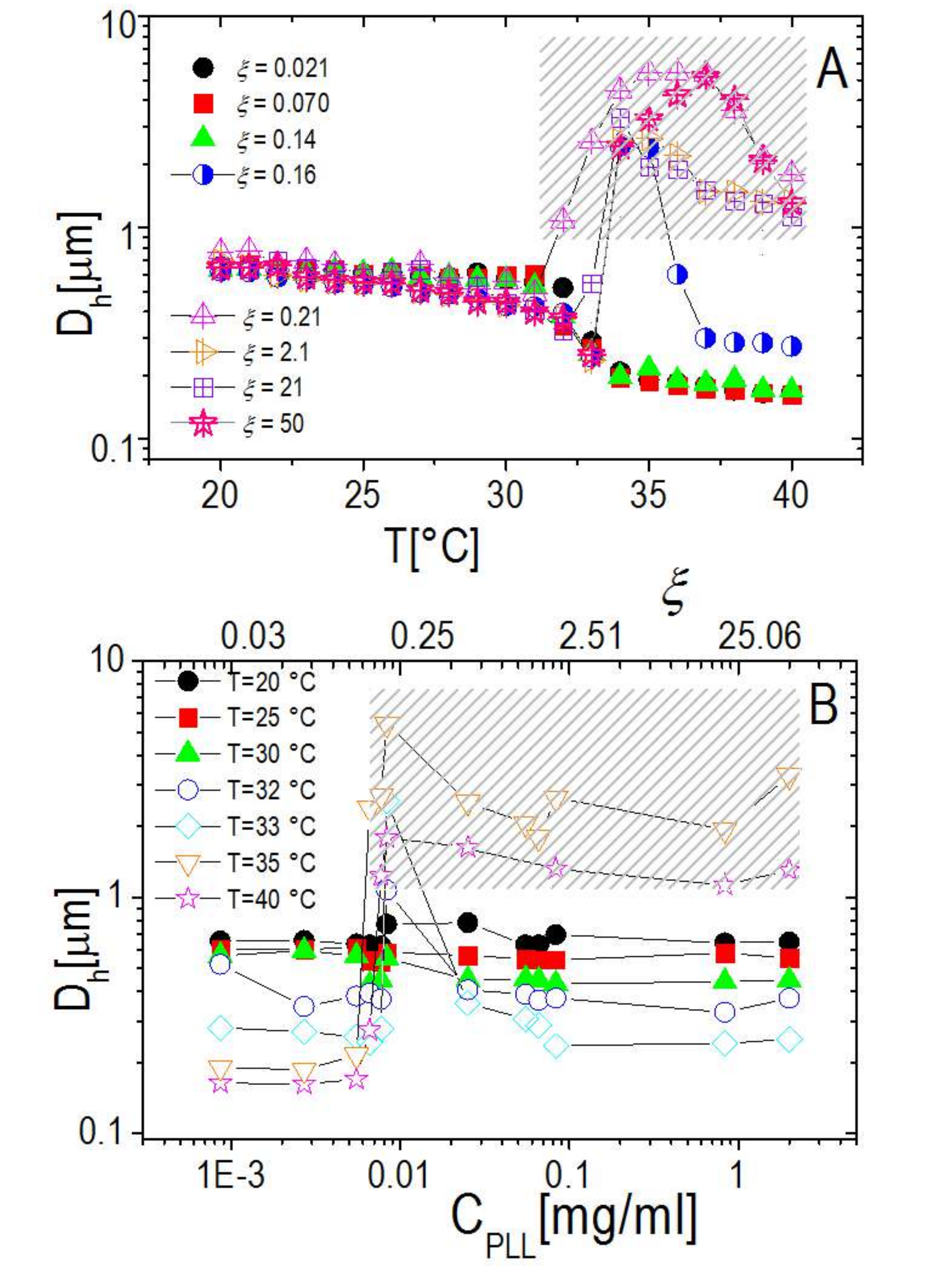}\\
  \caption{Panel A: Hydrodynamic diameter of PLL-PNiPAM complexes as a function of temperature for different PLL concentrations (charge ratios $\xi$) as indicated in the figure. Panel B: Hydrodynamic diameters of PNiPAM-PLL complexes in function of $C_{PLL}$ ($\xi$) for selected temperatures as indicated in the figure. The shaded regions in both panels enclose all the samples where flocculation has been observed after 12 hours.}\label{Diam-panel}
\end{figure}

Similarly to other colloid-polyelectrolyte systems, neutralization and overcharging of PNiPAM microgels is accompanied by clustering that depends on PNiPAM-PLL charge ratio and temperature, the latter being decisive, in this specific case, for the reentrance of the colloidal aggregation. Figure \ref{Diam-panel}-A shows the hydrodynamic diameters,  $D_h(T)$,  as a function of temperature for selected $\xi$ values. The microgel stability is substantially unaffected by polyelectrolyte addition for $C_{PLL}< 0.0066$ mg/ml ($\xi<0.16$): the measured hydrodynamic diameters follow the same critical behavior of the bare microgels. However, for $C_{PLL}= 0.0066$ mg/ml ($\xi$=0.16) the size of clusters shows unambiguously a maximum: the complexes form large aggregates only in the narrow range $T_c<T< T_{c\mu}$, while stable submicrometric clusters characterize the suspensions above the ET, where microgels deswell and become densely charged. As the concentration of $\epsilon-PLL$ is further increased, complex destabilization occurs at lower temperatures and the reentrance of the microgel condensation is suppressed. We interpret such finding as due to the high ionic strength of the suspensions for high $C_{PLL}$: as a matter of fact, for high $\epsilon-PLL$ concentrations, the PE counterions and the free polyelectrolyte chains contribute to screen the residual repulsion between complexes and one recovers the same phenomenology observed in the presence of monovalent salt (Figure \ref{NaCl-Panel}).

Interestingly the same reentrant behavior appears when $D_h$ is plotted vs $\xi$ for different temperatures (Figure \ref{Diam-panel}-B). Below the VPTT microgels do not significantly aggregate and, on the contrary, we observe a slight, albeit unambiguous, deswelling for $T=30$ $^{\circ}$C due to the screening of the microgel charges given by the $\epsilon$-PLL adsorption. For $T\approx T_c$ the typical reentrant condensation phenomenology appears: large micrometric aggregates form at $\xi=0.21$ and dissolve once a strong overcharging occurs for larger $\epsilon$-PLL concentrations. It's worth noting that the aggregation peak does not occur exactly where the mobility modulus shows a minimum for 32 $^{\circ}$C and 33 $^{\circ}$C (see figure \ref{mobility-CPLL}). Clustering occurs before a complete neutralization is attained.
This is not very surprising being the aggregation synergically driven by both charge heterogeneity and hydrophilicity of the PNiPAM-PLL complexes: charge heterogeneity is tuned by the PLL adsorption that screens hydrophobic interactions of near-critical microgels and it is not necessarily maximized at the charge-inversion point\cite{sennato_decorated_2012}, where $\mu \approx 0$; hydrophobic interactions are simply tuned by temperature. A more detailed study of the interplay between hydrophobic interactions and charge patch attraction at the VPT goes beyond the scope of this work and will be the subject of a future publication.
Finally, when $T$ is further increased large unstable clusters do not re-dissolve at large PLL content, as free polyion chains act as screening multivalent ions and give rise to the same phenomenology observed at large NaCl concentrations.
Therefore our findings point out an unprecedented and non trivial feature of thermosensitive polyion-microgel complexes: a reentrant condensation may occur by progressively adding oppositely charged polyions at fixed temperature or increasing temperature at fixed polyion content.
\begin{figure}[htbp]
  \includegraphics[width=7cm]{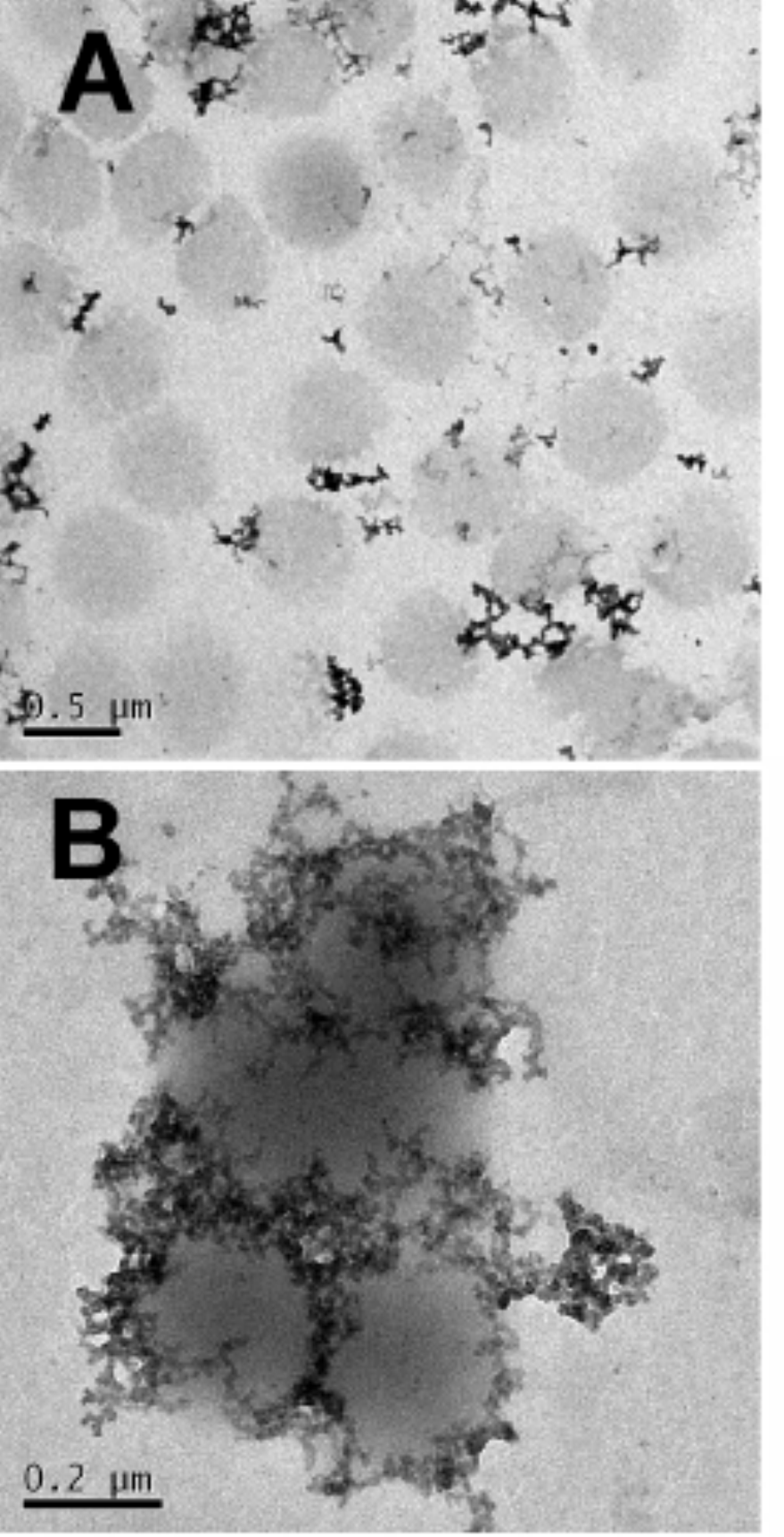}\\
  \caption{TEM images of PNiPAM-PLL sample prepared at $\xi=1.0$ at $T=25$ $^{\circ}$C (panel A), where individual swelled microgels are visible and PLL chains are free in solution. By heating up to 40 $^{\circ}$C (panel B) the aggregation of the PNiPAM-PLL microgel is promoted. Both images are obtained by PTA staining.}\label{TEM2}
\end{figure}
\begin{figure}[htbp]
  \includegraphics[width=7cm]{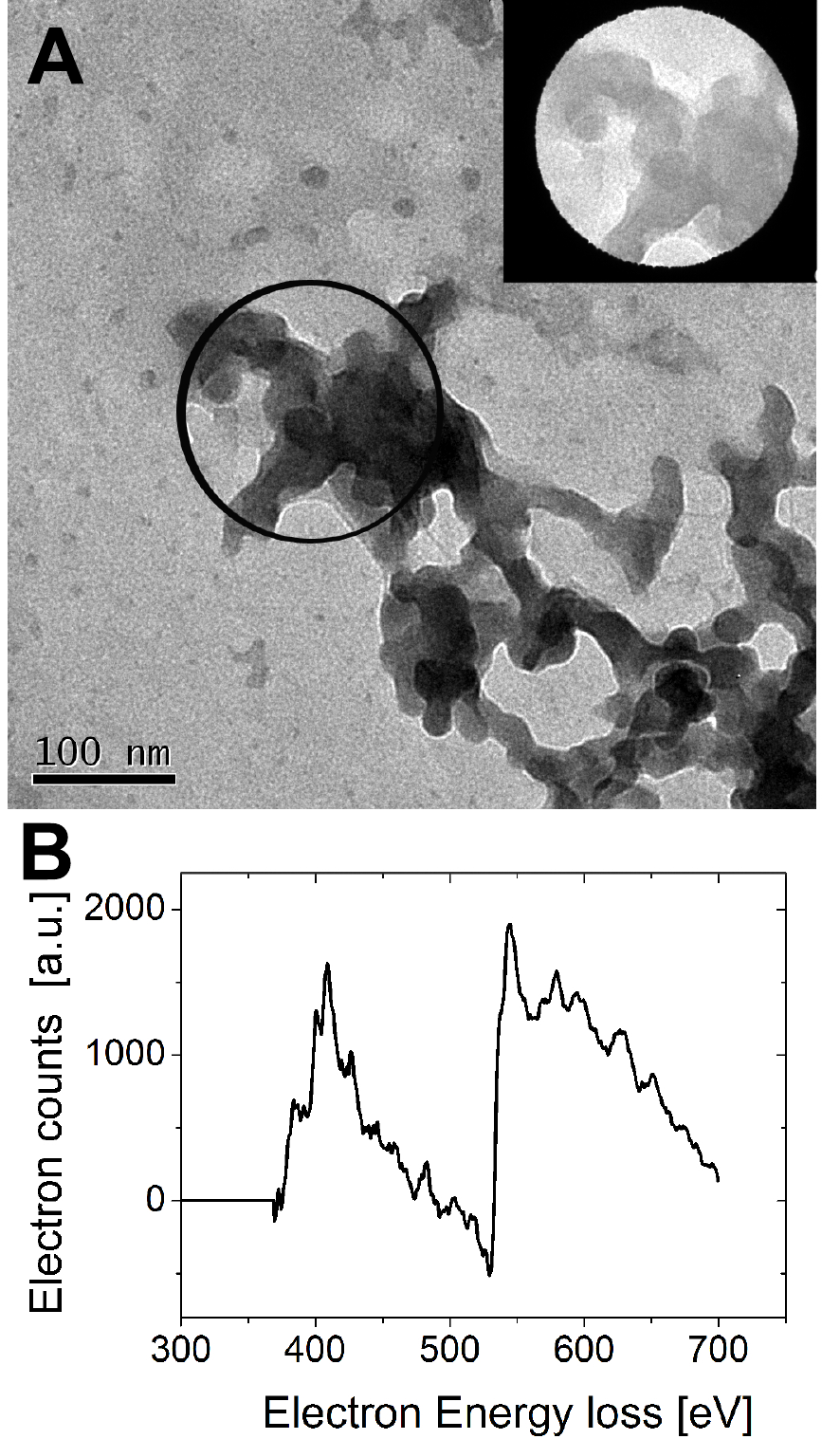}\\
  \caption{TEM image of a free PLL molecules resembling a disordered knot obtained by PTA staining (panel A). By focusing the electron beam in the marked circular region of the knot, EELS has been performed to detect the presence of nitrogen and oxygen, without damaging the sample, as the post-EELS image reported in the inset testifies. The EELS spectrum of inner-shell ionization electrons is shown in panel B. After background subtraction the characteristic nitrogen and oxygen edge peaks are detected at 401 and 532 mV respectively, thus confirming that the visualized knot contains $\epsilon$-PLL chains.}\label{TEM3}
\end{figure}
In order to confirm the overall emerging scenario we have performed TEM measurements on selected mixtures, in a range of $\xi$ where clustering is observed. Figure \ref{TEM2} shows two images at $\xi=1.0$, below (panel A) and above (panel B) the VPT. Due to the PTA staining, microgel particles appear light grey  while the positively charged $\epsilon$-PLL chains appear as darker knots since they are able to attract the negatively charged PTA. Crossing the VPT, the $\epsilon$-PLL chains unambiguously pass from being free or poorly adsorbed to an adsorbed state promoting microgel aggregation in a glue-like fashion, where $\epsilon$-PLL chains preferentially occupy the interstitial regions: $\epsilon$-PLL chains act as an electrostatic glue. We have further checked the validity of this assessment by performing transmission Electron Energy Loss Spectroscopy (EELS) by
gathering the electrons transmitted from a circular portion of the sample  occupied only by the dark spots emerging in all images containing PLL chains, to ensure their correct attribution to the polyelectrolyte. The result of EELS experiment is shown in figure \ref{TEM3}, where panel A shows the PLL knot with the internal circular portion where the corresponding EELS spectrum shown in panel B has been determined. It has to be noted that the sample is not damaged by the EELS experiment, as shown in the inset of panel A. To get the contribution of the sample, we removed the background by fitting with a power law the collected data in an energy window extending before the nitrogen edge. The obtained spectrum is characterized by two peaks at 401 eV and 532 eV corresponding to the nitrogen K-edge and oxygen K-edge, respectively. Being the sample stained by PTA, the peak of oxygen contains also the contribution of the staining molecules, while the presence of Nitrogen unambiguously testifies the presence of PLL in the knot.

\subsubsection{Dielectric Spectroscopy}
\begin{figure}[htbp]
\begin{center}
 \includegraphics[width=10cm]{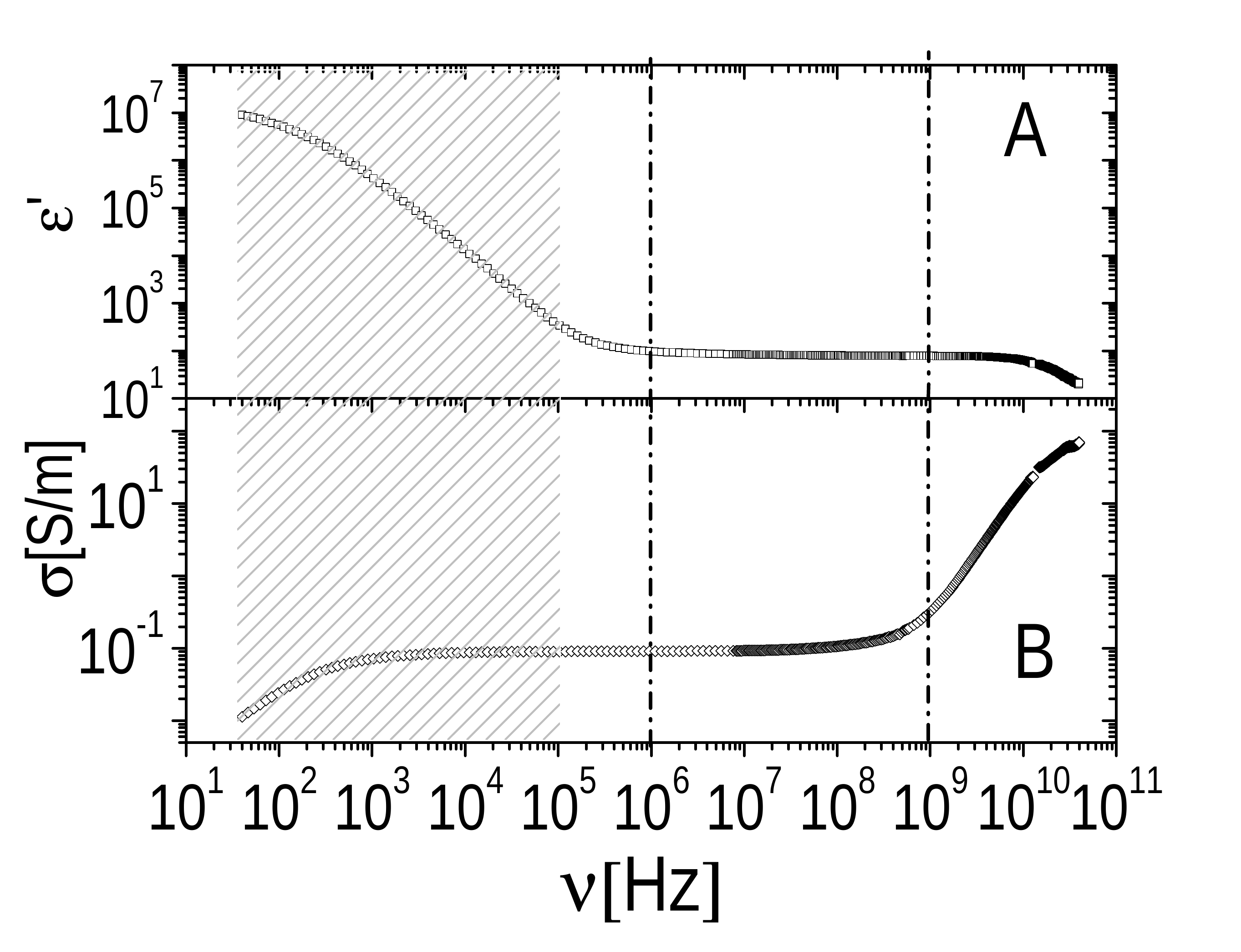}
\caption{Typical real part of the complex permittivity $\varepsilon'$ (Panel A) and conductivity  $\sigma$ (Panel B) measured for PNIPAM-PLL samples (here $\xi=0.5$, $\varphi=0.53$ and T=26 $^{\circ}$C). In the low frequency tail of the spectrum (shaded region) the large increase of $\varepsilon'$ as well as the strong decrease of $\sigma$ are due to the electrode polarization of the measuring cell. This contribution is subtracted before data analysis as described in section \ref{dielectric}. The two vertical dash-dotted lines delimit the region of interest of the spectrum that has been further analyzed as discussed in the main text.}
\label{SpectraRaw}
\end{center}
\end{figure}

\begin{figure}[htbp]
\begin{center}
 \includegraphics[width=10cm]{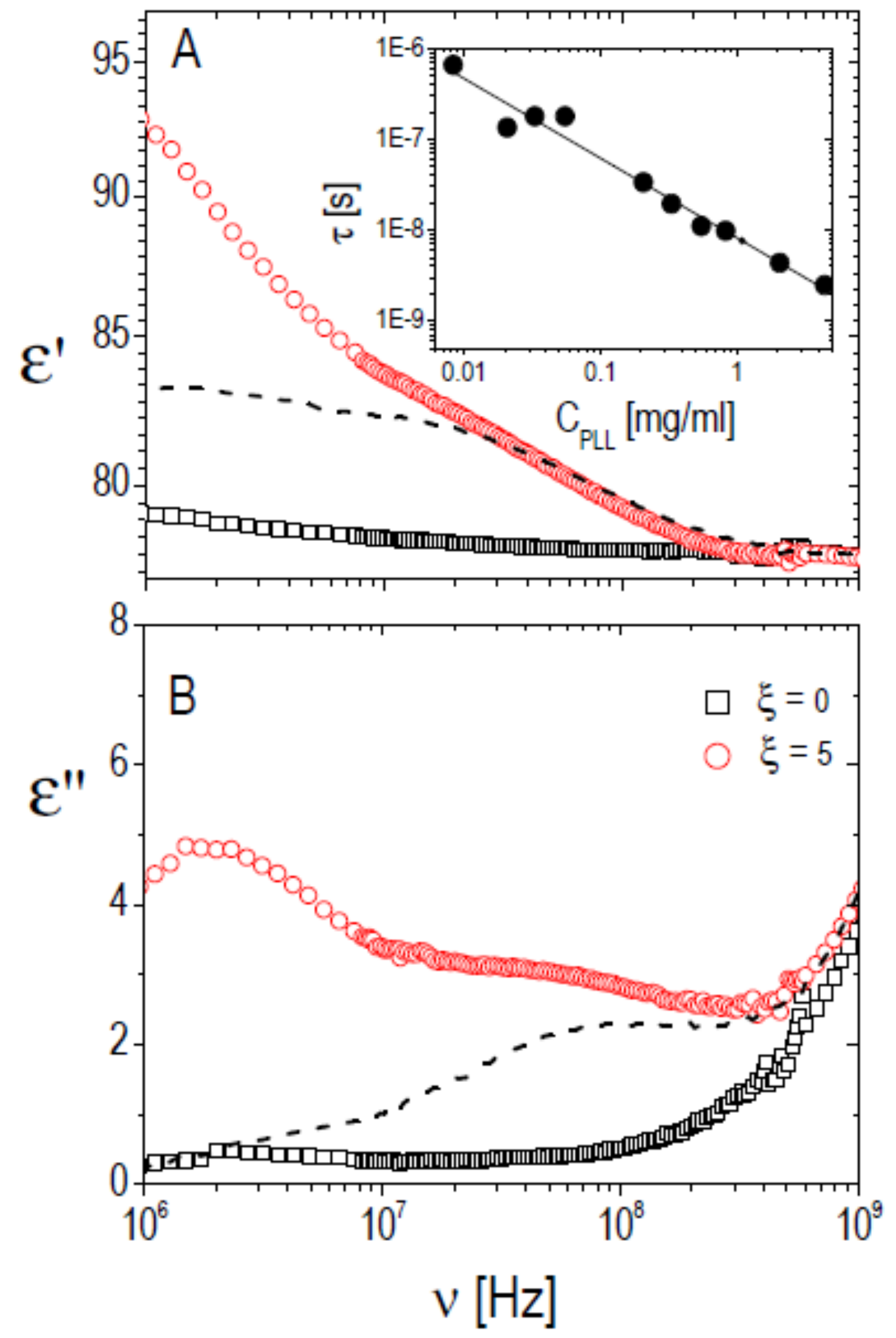}
\caption{
Typical behavior of the real ($\varepsilon'$, panel A) and imaginary part ($\varepsilon''$, panel B) of dielectric spectra of PNiPAM-PLL suspensions (full lines) measured at 26 $^\circ$C, with (circles) and without (squares) the addition of PLL at $\varphi=0.53$. The spectra of pure PLL suspensions is also shown (dashed lines). As can be seen, the lower part of the spectrum is markedly different in the two cases.
Inset: relaxation time of the 'intermediate' polyelectrolyte relaxation of pure $\epsilon$-PLL samples as a function of $C_{PLL}$. In the $C_{PLL}$ range considered for our DS experiments a dependence $\tau\sim C_{PLL}^{-1}$ (solid line) is observed.
}
\label{Fig:Spectra}
\end{center}
\end{figure}
We have further investigated the adsorption of $\epsilon$-PLL on PNiPAM microgels by dielectric spectroscopy. Figures \ref{SpectraRaw} and \ref{Fig:Spectra} show typical dielectric spectra below the microgel VPT (T=26 $^\circ$C). The first shows a representative raw dielectric spectrum ($\varepsilon'$ and $\sigma$), including the electrode polarization effect in the whole frequency range accessible to our experiments; the second shows only the portion of the spectrum in the enclosed frequency range 10$^5$-10$^9$ Hz measured for a PNiPAM-PLL mixture ($\xi$=5) after having subtracted the electrode polarization contribution as discussed in section \ref{dielectric}.
In figure \ref{Fig:Spectra} the spectra of pure $\epsilon$-PLL and PNiPAM aqueous solutions are also shown for comparison. 
In the high frequency wing, the small increase visible in $\varepsilon''$ and the corresponding decrease (barely visible on this scale) in $\varepsilon'$ are due to the onset of the solvent (water) relaxation, centered at $\approx$ 20 GHz \cite{ellison_water:_1996-1}(see also figure \ref{SpectraRaw}). As expected, in this frequency range, due to the very low mass concentration of the polymer and to its low intrinsic polarizability, the spectra of pure PNiPAM microgel aqueous solutions ($\xi=0$) appear almost flat, except for the water contribution.

The relaxation centered slightly below 100 MHz is present both in pure $\epsilon$-PLL and in microgel-PLL suspensions, while it is not present in absence of polyelectrolyte ($\xi=0$).
This 'intermediate frequency relaxation', due to counterion fluctuation, is typical of polyelectrolyte solutions and is characterized by a power law dependence of the relaxation time on polyelectrolyte concentration \cite{bordi_04} in pure $\epsilon$-PLL samples, as shown in the inset of figure \ref{Fig:Spectra} (Panel A).

\begin{figure}[htbp]
\begin{center}
 \includegraphics[width=15cm]{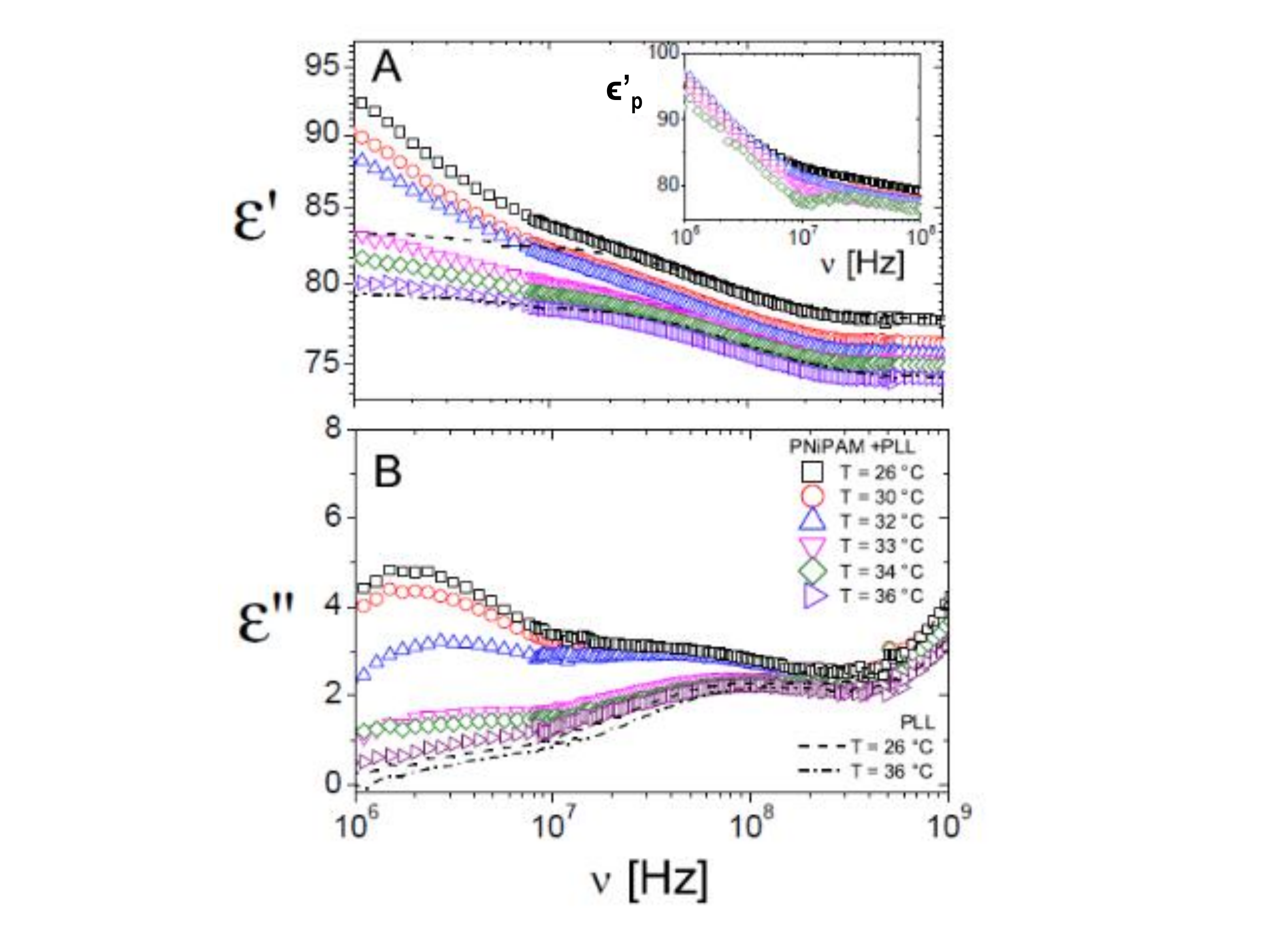}
\caption{Typical behavior of the real ($\varepsilon'$, panel A) and imaginary part ($\varepsilon''$, panel B) of dielectric spectra of PNiPAM-PLL suspensions measured at different temperatures below and above the VPT, for $\xi=5$ and $\varphi$(20 $^{\circ}$C)=$0.56$. For comparison, the spectra of pure PLL suspensions at the lowest and highest temperatures considered, 26 $^\circ$C and 36 $^\circ$C, are also shown  (dashed and dot-dashed lines, respectively). Inset shows the frequency dependence of the real part of the particles permittivity, $\epsilon'_p$, calculated from the corresponding measured spectra using the Looyenga model (eq. \ref{Looyenga_eq}). The amplitude of the dispersion in the lower frequency range, which is associated with the presence of the PNiPAM-PLL complexes, significantly decreases across the transition, due to the particle deswelling and the consequent decrease of their volume fraction. However, the calculated particles permittivity (inset) does not change appreciably across the transition, maintaining its frequency dependence.}
\label{Fig:SpectraT}
\end{center}
\end{figure}

However, in the spectra of all the PNiPAM-PLL samples, in addition to the 'intermediate frequency relaxation' due to the non-adsorbed PLL chains, a rather pronounced relaxation appears in the low frequency wing (see figure \ref{Fig:SpectraT}). Based on the structural information obtained from light scattering and TEM images, we can attribute this relaxation to the presence of a shell around the microgel particles, formed by the adsorbed polyelectrolyte chains. In fact, the amplitude of this dispersion increases approximately in proportion to the microgel concentration. However, what is perhaps more interesting in the present context, is that this amplitude shows a strong dependence on temperature, decreasing significantly across the volume phase transition.\\

On the basis of the above considerations, all spectra have been fitted with a complex function containing three relaxations: i) a Debye relaxation occurring at $\sim$20 GHz due to local fast rearrangements of water molecules, whose parameters (relaxation time and dielectric increment) are tabulated in literature\cite{ellison_water:_1996-1} and ii) two relaxations modeled by two complex Cole-Cole equations\cite{kalmykov_microscopic_2004}. The latter, as just mentioned, are attributed to the $\epsilon$-PLL counterion relaxation and to the onset of a dielectric discontinuity given by the formation of the polyelectrolyte shell on the microgel periphery.

Once the parameters describing these three dispersions are obtained, we proceed as follows: we assume that the PNiPAM-PLL suspensions can be described as homogeneous mixtures of isotropic particles, with complex permittivity $\tilde{\epsilon}_p$, uniformly dispersed in a continuous medium with complex permittivity $\tilde{\epsilon}_m$ at a volume fraction $\varphi$. We then use the Looyenga equation \ref{Looyenga_eq} to calculate an "effective permittivity" of the suspended particles $\tilde{\epsilon}_p$ from the measured total permittivity of the suspension $\varepsilon$. To this end, we assume that $\tilde{\epsilon}_m$ is given by the sum of the two relaxations observed in the high frequency wing of the spectrum, due to water and free PLL, as described above. The value of $\varphi$ for all temperature has been obtained according to $\varphi(T)=\varphi(20 ^{\circ}C)\frac{D_h^3(T)}{D_h^3(20 ^{\circ}C)}$ , where $\varphi(20 ^{\circ}C)$ has been obtained via viscosimetry as discussed in section \ref{visco}. The dc conductivity of the solvent $\sigma_m$ is left as an adjustable parameter, and it has been determined by requiring that i) the MHz range of the resulting $\tilde{\epsilon}_p$ is either flat or described by a Maxwell-Wagner-Sillars (MWS) dispersion and ii) smoothly converges to $\varepsilon$ in the high frequency limit.


Indeed were the particles dielectrically homogeneous, their effective permittivity would be $\tilde\epsilon_p=\epsilon'_p+i\omega \epsilon_0 \sigma_p$, with $\epsilon'_p$ and $\sigma_p$ independent of the frequency.
Conversely, particles presenting internal dielectric discontinuities or interfaces would show a frequency dependent effective permittivity. This is the well known Maxwell-Wagner effect \cite{DielectricSpectroscopy03}.\\
The inset of figure \ref{Fig:SpectraT} shows the effective permittivity of the decorated microgel particles, $\tilde\epsilon_p$ , calculated from the measured dielectric spectra, at different temperatures across the VPT. A strong dependence on the frequency, that is the signature of the presence of a dielectric discontinuity, is observed at all temperatures. However, what is even more noticeable, is that although the amplitude of the dispersion neatly decreases across the transition, the effective particle permittivity calculated from this dispersion is scarcely affected by temperature. Interestingly the curves of  $\tilde\epsilon_p$ vs $\nu$ calculated at the different temperatures almost superimpose, even though, due to the decreasing amplitude of the dispersion the calculated values becomes increasingly scattered. This behavior suggests a substantial invariance of the shelled structure of the particles across the transition, with the observed neat decrease of the amplitude of the relaxation mainly due to the strong decrease of the shelled particles volume fraction due to their shrinkage.

\begin{figure}[htbp]
\begin{center}
\includegraphics[width=10cm]{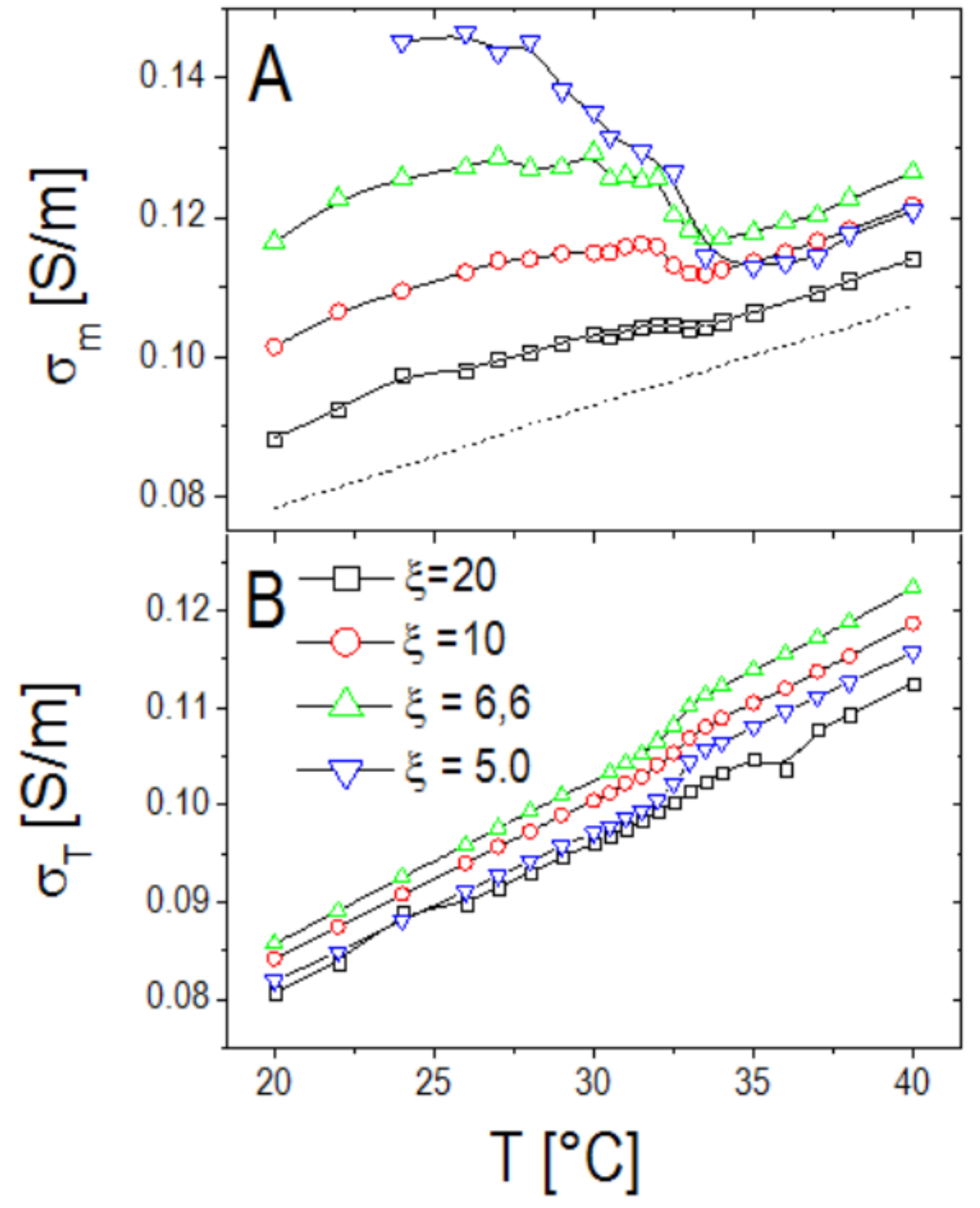}
\caption{
Panel A: Solvent conductivity as obtained from equation \ref{Looyenga_eq} through the procedure described in the text for different charge ratio $\xi$. For comparison, the conductivity of pure PLL suspensions at the same $C_{PLL}$ as the PNiPAM-PLL mixtures ($C_{PLL}=$4.4 mg/ml) is also shown (dashed line).
The charge ratio has been tuned here by varying the PNiPAM concentration from $\varphi$(20$^{\circ}$C)=$0.56$ down to $\varphi$(20$^{\circ}$C)=$0.14$.
Panel B: Total conductivity of the PNiPAM-PLL solutions $\sigma_T$ as a function of temperature, at the same charge ratios shown in panel A.
}
\label{SigmaAllT}
\end{center}
\end{figure}

The static solvent conductivity $\sigma_m$ that we obtain from this analysis shows an interesting behavior (Figure \ref{SigmaAllT}, panel A). While the measured dc conductivity of the suspensions $\sigma_T$ shows a small but significant increment at the VPT (Figure \ref{SigmaAllT}, panel B), $\sigma_m$ significantly decreases with temperature. At the same time the conductivity of the decorated particles remains almost constant as we can infer from the invariance of the real part of $\tilde{\epsilon}_p$ (inset of figure \ref{Fig:SpectraT}).
Notwithstanding the limitations of the procedure, the unambiguous decrease of $\sigma_m$ suggests that, due to a large volume of water expelled from the microgels, the net ionic strength of the solvent decreases because of the dilution of the 'ionic atmosphere' around the decorated particles when the VPT is crossed and conforms to a scenario where the observed increase of the total conductivity is mainly due to the large decrease of the microgel volume fraction, and not to a significant expulsion of counterions from the inner part of the microgels.

\subsection{Thermal reversibility}\label{revtherm}
\begin{figure}[htbp]
  \includegraphics[width=12cm]{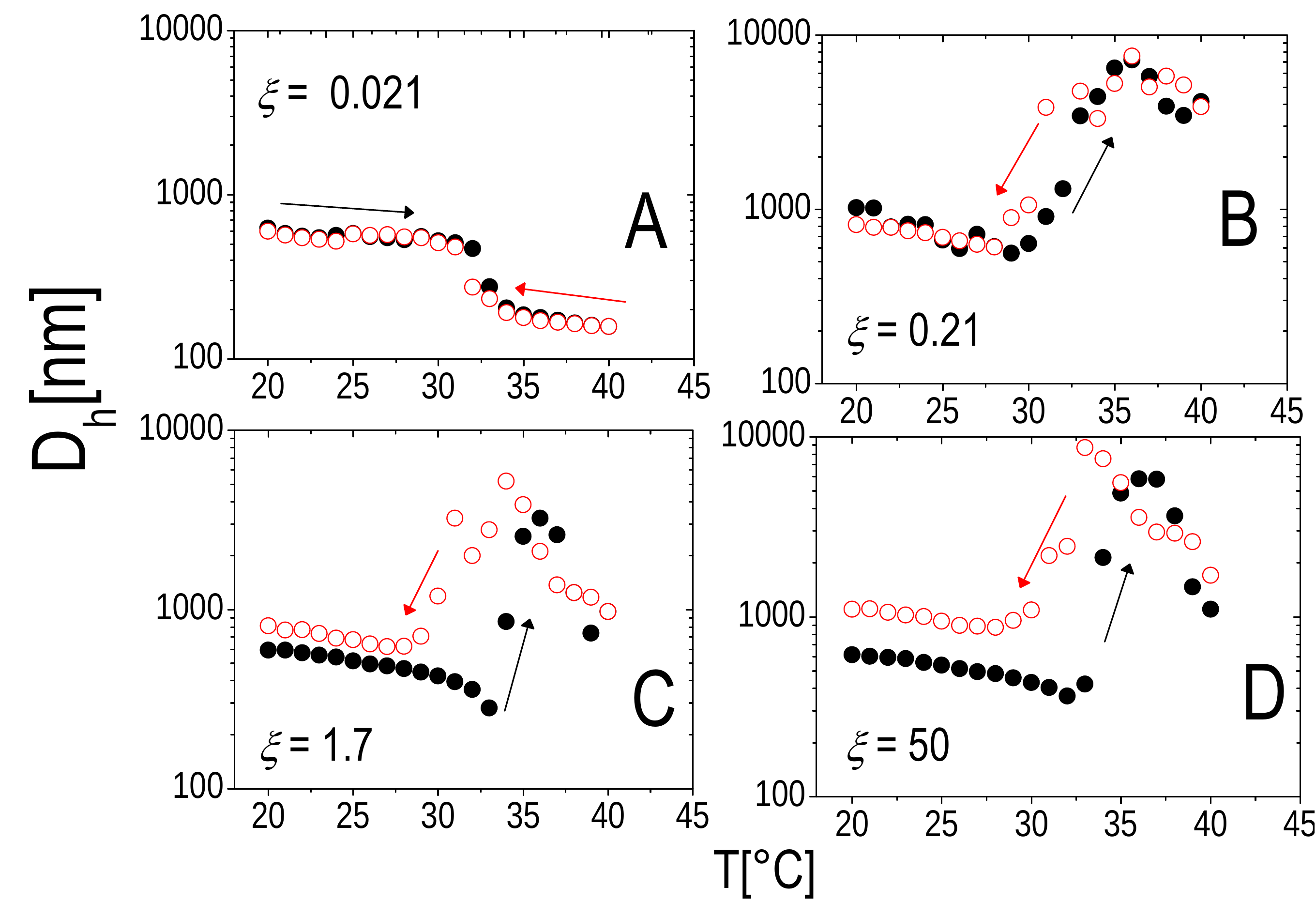}\\
  \caption{Hydrodynamic diameter $D_h$ in function of temperature during heating (black full circles) and cooling (empty red circles) ramps with thermalization time $t_{therm}=300$ s for different charge rations $\xi$ as indicated in the panels.}\label{reversibility}
\end{figure}

To test the thermal reversibility of the self-assembly of microgels decorated by $\epsilon$-PLL,  we have performed temperatures cycles for all the polyion concentrations according to the following thermal protocol:  i) a first ascending ramp from 20 $^{\circ}$C to 40 $^{\circ}$C by increasing temperature of 1 $^{\circ}$C each time. Before each measurement the samples have been left to thermalize 300 s at the target temperature (standard protocol already described in section \ref{size-zeta meas}). ii) a descending ramp from 40 $^{\circ}$C down to 20 $^{\circ}$C has then been carried out with the same temperature step and thermalization time of i). The results are shown in figure \ref{reversibility} for selected concentrations of polyelectrolyte.
At low concentration of $\epsilon$-PLL the stability of the microgel suspension is not affected at all and no aggregation nor thermal hysteresis is observed (panel A): the charge heterogeneity introduced by the polyelectrolyte adsorption does not give rise to enough attraction to compensate the electrostatic repulsion between microgels. This occurs at any temperature of the thermal cycle. An increase of polyelectrolyte concentration (panel B) induces the formation of large, finite size clusters. This aggregation is reversible and there is no appreciable hysteresis. However, as the PLL concentration is further increased to  $\xi\approx 1.6$ (panel C) a significant hysteresis in the aggregation behavior appears,  the dissolution of the clusters occurring at about 5 $^{\circ}$C below the VPT. A large increase of PLL concentration (panel D) does not changes qualitatively this behavior but for an increase of the size of the residual clusters.

We interpret the presence of thermal hysteresis as the signature of the large asymmetry of time scales characterizing the adsorption of the polyelectrolyte on the external shell of the microgels and the cluster dissolution. The first is driven by both the polyion and the single microgel diffusion and gives rise to the (almost instantaneous) aggregation of decorated microgels as temperature is raised above $T_{c\mu}$. This, i.e. the rapid cluster formation as polyelectrolytes are mixed with oppositely charged colloids, has been observed in all polyelectrolyte-colloid mixtures and has been discussed within the framework of a kinetically arrested (metastable) clustering \cite{bordi_polyelectrolyte-induced_2009,sennato_decorated_2012,sennato_salt-induced_2016}.
On the contrary large cluster dissolution is driven by both the time scale associated to the polymer desorption from the microgel and the one associated to microgel intra-cluster diffusion after cooling the system. The latter, being dominated by the complete disentanglement of aggregated microgels, is a much larger time scale than the former. Moreover, close-packed microgels must be thought as weakly interpenetrating particles \cite{mohanty_interpenetration_2017, wolfe_rheology_1989} whose dynamics is necessarily affected by multiple contacts between the dangling chains present at their periphery.

The cluster dissolution (partial or total) observed by lowering the temperature is a direct evidence that polyelectrolyte adsorption is a reversible process: polyelectrolytes desorption occurs as microgels re-swell and their charge density decreases. However the complete dissolution does not occur within the time of our experiment when the polyelectrolyte content is high. Micrometric and submicrometric clusters do not dissolve and are stable in solution for several hours, their size being different from that of the initial microgels by an amount that increases for increasing polyelectrolyte content. This corroborates the idea that complete particle dissolution is dictated by a much larger time scale.

On the other side, the mobility of the decorated microgels does not show significant hysteresis under thermal cycles. This conforms to what has been said in section \ref{salt}: $\mu$ is univocally determined by the charge density and the friction coefficient of the suspended objects, two intensive quantities that, being fixed by temperature, are not sensibly affected by clustering and the thermal history of the suspensions.

\begin{figure}[htbp]
  \includegraphics[width=12cm]{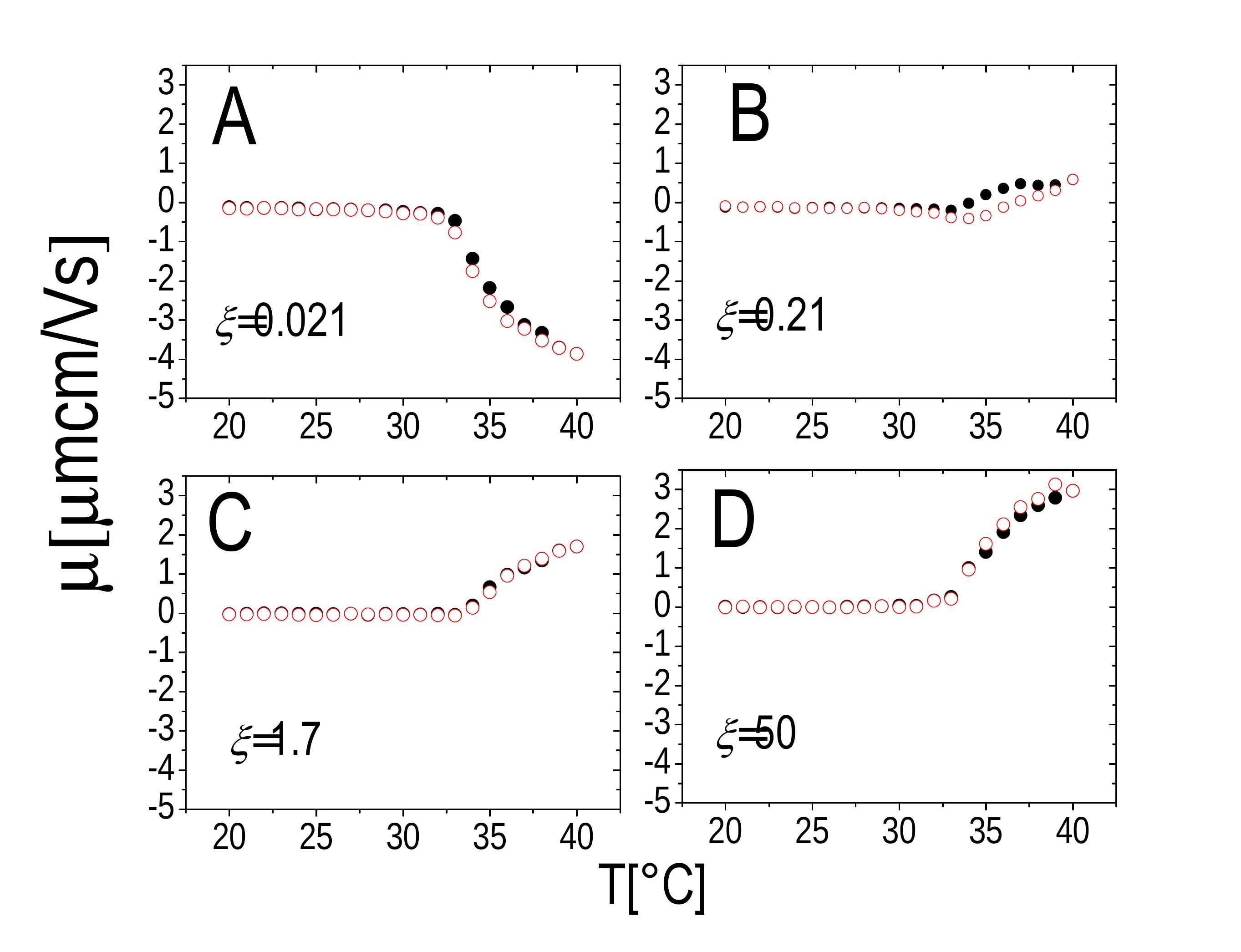}\\
  \caption{Mobility $\mu$ as a  function of temperature during heating (black full circles) and cooling (empty red circles) ramps with thermalization time $t_{therm}=300$ s for different charge rations $\xi$ as indicated in the panels.}\label{reversibility_1}
\end{figure}

The stability of PLL-microgel suspensions will not be discussed further here and will be the subject of a future work.

\subsection{Effect of monovalent salt on microgel-polyelectrolyte complexation}
\begin{figure}[htbp]
  \includegraphics[width=12cm]{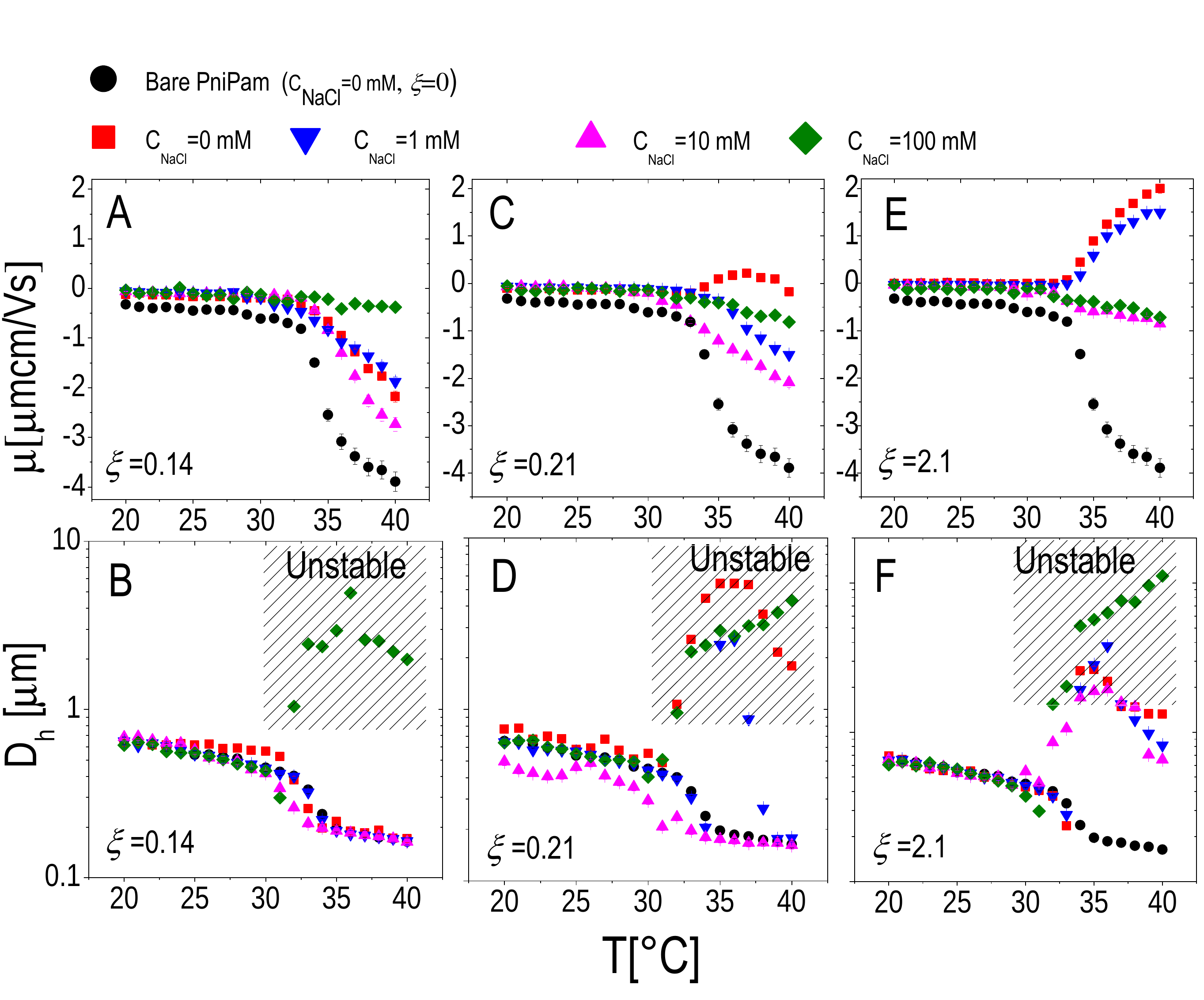}\\
  \caption{Electrophoretic mobility $\mu$ (panels A, C, E) and hydrodynamic diameters $D$ (panels B, D, F) in function of the temperature for selected salt concentrations and charge ratios as indicated in the figure.}\label{salteffect}
\end{figure}

Finally we have tested the effect of the addition of monovalent salt (NaCl) on the complexation between PNiPAM microgels and $\epsilon$-PLL chains by measuring the electrophoretic mobility and the size of the suspended particles as a function of temperature for several salt concentration.
In figure \ref{salteffect} we show the results obtained for selected charge ratios to point out the effect of the screening for different amounts of adsorbed polyelectrolyte. We still distinguish two regimes: the subcritical swollen microgels for $T<T_c$ and the shrunk densely charged microgels for $T>T_c$. Below the VPT microgels are poorly 'decorated' by the PLL layer and not densely charged. This results in a very weak dependence of both $\mu$ and $D_h$ on salt concentration, both being very close to the values obtained with no added salt.
The scenario radically changes above $T_c$, the suspended microgels being more densely charged, more densely covered by electrostatically adsorbed PLL chains and hence more affected by a drastic change of the ionic strength. In particular for low $\xi$ (Figure \ref{salteffect}-A/B), where the overcharging does not occur, the addition of salt does not induce any change of the mobility sign as expected, although a non monotonic behavior is visible at high temperatures: $\mu$ first decreases, becoming more negative, and then goes up to zero due to the high screening. Consistently, the size of particles is also unaffected by the addition of salt but for very high salt content ($C_{NaCl}>$30mM), where suspensions are destabilized and flocculation occurs.
As the charge ratio increases, microgels get overcharged above $T_c$ (Figure \ref{salteffect}-C/D/E/F), large variations of mobility are enhanced and we clearly observe a change of the sign of $\mu$ that, before converging to zero due to the high screening, passes from positive to negative values by increasing the salt content, signaling the desorption of the PLL chains from the microgels. This is indeed expected for the screening-reduced regime in polyelectrolyte electroadsorption \cite{muthukumar_adsorption_1987,VandeSteeg1992,forsman_polyelectrolyte_2012} that has been observed in numerous cases in PE-colloid mixtures \cite{rojas_effect_1998,sennato_salt-induced_2016} and in simulations \cite{truzzolillo_interaction_2010,forsman_polyelectrolyte_2012,de_carvalho_first-orderlike_2010,ulrich_many_2006}.
It is worth here to describe more in detail the aggregation and distinguish the cases of weakly and strongly overcharged microgels. In figure \ref{salteffect}-C/D we show how PLL-microgels complexes at $\xi=0.21$ and $T>T_c$ pass from a weakly overcharged state ($C_{NaCl}=0$ mM), characterized by large unstable clusters at high temperatures to non-overcharged states at $C_{NaCl}=1$ mM and $C_{NaCl}=10$ mM. For these two salt concentrations the temperature dependence of the aggregate size is particularly interesting: for $C_{NaCl}=$1 mM the PLL-microgels aggregate in a relatively narrow range of temperatures $35 ^{\circ}$C$\leq T\leq 37 ^{\circ}$C eventually dissolving and forming stable single decorated microgels for $T>37$ $^{\circ}$C, where their large charge density and the enhanced screening prevent aggregation and massive PLL adsorption respectively; at $C_{NaCl}=10$ mM PLL desorption is even more pronounced and microgels do not aggregate in the entire range of temperature investigated. This is an indirect demonstration that the microgel aggregation is indeed induced by PLL adsorption and not trivially by the increased ionic strength and shows unambiguously that PLL adsorption is dominated by electrostatics rather than more specific affinities between lysines and NiPAM monomers.\\
Finally, as the size ratio is increased up to $\xi=2.1$ (Figure \ref{salteffect}-E/F), microgels get highly overcharged, unstable clusters are observed at all salt concentrations for $T>T_c$ while a change of the mobility sign is again observed and confirms the PLL desorption.

\section{Conclusions}
We have investigated the complexation of thermoresponsive ionic microgels with oppositely charged polyions. We have shown that microgel overcharging is triggered by their volume phase transition tuning the charge density of the particles. Collapsed microgels are able to adsorb a large fraction of suspended polyions and this adsorption causes a "multivariable" reentrant condensation: at fixed polyion concentration clustering occurs at the microgel VPT and may disappear once the temperature is further increased due to the large polyion adsorption; similarly, at fixed temperature, aggregation is triggered only by polyion adsorption and shows a reentrant behavior for near-critical microgels. \emph{This phenomenon is new} and it will be further investigated by using different polyelectrolytes to test the role played by their molecular weight and hydrophobicity.
Besides the electrophoretic and the size characterization of the complexes, we have probed the aforementioned phenomenology by means of TEM and DS experiments that confirmed polyion adsorption, the consequent formation of a dielectric discontinuity at the microgel periphery and the glue-like behavior of the adsorbed chains. We have tested thermal reversibility showing that, on the time scale of our experiments, clustering is quasi-reversible: complete cluster dissolution is not completely attained after one thermal cycle probably due to a larger time scale characterizing microgels' disentanglement within a cluster at high polyion content, while electrophoretic mobility does not depend on the thermal history of the mixtures as expected for polymer-based particles. Finally, by probing the polyion-induced microgel aggregation at different uni-univalent salt content we have shown that polyion adsorption is electrostatic in nature and that desorption may occur once one crosses a salt concentration threshold.
By showing that the VPT of thermoresponsive ionic microgels can be employed to trigger polyion adsorption and tune reentrant microgel condensation, our work lays the foundation for a groundbreaking strategy to tune electroadsorption ruled by temperature that can be employed in a variety of fields spanning wastewater and soil remediation, nanoencapsulation of small charged nanoparticles, and selective drug delivery.

\begin{acknowledgement}
 D.T. acknowledges the The Young Investigator Training Program (YITP) financed by ACRI (Italian Banking Foundation Association) in association with the European Colloid and Interface Society (ECIS) Conference 2017. S.S. acknowledges E. Zaccarelli for valuable discussions and support from the European Research Council (ERC Consolidator Grant 681597, MIMIC)
\end{acknowledgement}




\providecommand{\latin}[1]{#1}
\providecommand*\mcitethebibliography{\thebibliography}
\csname @ifundefined\endcsname{endmcitethebibliography}
  {\let\endmcitethebibliography\endthebibliography}{}

\end{document}